\documentclass[%
 preprint, %linenumbers,
 amsmath,amssymb,
 aps, physrev,
]{revtex4-2}

\usepackage{graphicx}% Include figure files
\usepackage{dcolumn}% Align table columns on decimal point
\usepackage{bm}% bold math
\usepackage[final]{changes}
\begin{document}

\preprint{APS/123-QED}

\title{\textbf{Two-stage dispersion mechanism of clean spherical bubbles rising in a chain} 
}% 

\author{Satoi Suzuki}
 %\altaffiliation[Also at ]{Department of Mechanical Engineering, Shizuoka University.}%Lines break automatically or can be forced with \\
\author{Toshiyuki Sanada}%
 \email{Contact author  : sanada.toshiyuki@shizuoka.ac.jp}
\affiliation{%
  Department of Mechanical Engineering, Shizuoka University, Chuo-ku, Shizuoka 432-8561, Japan
}%

%  \homepage{http://www.Second.institution.edu/~Charlie.Author}
% \affiliation{
%  First affiliation for this author
% }%
% \affiliation{
%  second institution for this author
% \affiliation{%
%  Authors' institution and/or address\\
%  This line break forced with \textbackslash\textbackslash
% }%


\date{\today}% It is always \today, today,

\begin{abstract}
Wake-induced lift is a key mechanism governing the initial destabilization of bubbles rising in a chain (Atasi \textit{et al.}, 2023). 
Moore's wake model predicts limited interfacial vorticity and a relatively slender, spatially confined wake for clean spherical bubbles, suggesting that wake-mediated interactions weaken as the inter-bubble spacing increases.
However, we observed pronounced large-scale lateral dispersion and strong bubble frequency dependence in controlled experiments where bubble diameter and generation frequency were independently varied, even when the inter-bubble separation exceed the characteristic wake length. 
A reduced-order model incorporating pairwise wake-induced interactions captured the onset of bubble chain destabilization but systematically underpredicted the subsequent emergence of large-scale dispersion. 
We demonstrate that bubbles rising in a chain collectively generate a mean upward liquid flow that modifies the local shear field, enhancing the lateral migration through shear-induced lift. 
Incorporating this self-induced weak flow into the model quantitatively reproduced both the dispersion magnitude and its frequency dependence.
These results suggest that the dispersion of bubbles rising in a chain involves a two-stage mechanism, with initial chain destabilization mediated by wake interactions, followed by flow modification arising from two-way coupling between bubbles and the liquid. 
This collective mechanism highlights the importance of self-induced mean flow effects in continuum descriptions of bubble flows.
\end{abstract}

\maketitle

%\tableofcontents

\section{\label{sec:level1}Introduction}
Bubbles rising in a chain can exhibit a range of behaviors; for example, bubbles rising in champagne maintain stable straight-line trajectories, whereas  the trajectories tend to disperse in carbonated beverages. 
Previous studies have numerically and experimentally investigated the instability and dispersion of bubbles rising in a chain, with a focus on interactions between bubbles. 
Ruzicka \cite{Ruzicka2000}  incorporated hydrodynamic interactions into a model for the dynamic behavior of equally sized spherical bubbles, fixed in a vertical line, with $50\le Re\le 200$. 
The model predicted  that free-end bubble chains preferentially split into smaller bubble groups, while fixed-end chains behave in an unstable and chaotic manner. 
Sanada \textit{et al.} \cite{Sanada2005} reported experimental observations of the motion of a single nitrogen bubble chain at $300\le Re\le 650$. 
By independently controlling the bubble diameter and generation frequency, they identified that bubbles consistently followed similar paths at low frequency, but diverged and scattered at high frequency. 
This observation held even for regimes where an isolated bubble rose in a straight line. 
Atasi \textit{et al.} \cite{Atasi2023} classified stable and unstable regimes for bubble chains based on the bubble deformation and surface contamination. 
They also reported that the primary factor driving instability and dispersion is the lift force induced by the wake of the leading bubble. 
\added{While spherical bubbles experience positive lift, i.e., driving them toward higher relative velocity regions, deformed or contaminated bubbles can experience a lift force reversal. 
  Legendre and Zenit \cite{legendre2025gas} reviewed bubble chain stabilization while focusing on lift reversal. 
  They explained that bubbles with negative lift coefficients remain directly beneath the preceding bubble, where the relative velocity is lower, thereby establishing stable bubble chains.
  }
However, the relationship between wake-induced lift and bubble dispersion was not sufficiently quantitatively discussed in the regime where bubble deformation is negligible due to the difficulty of controlling the bubble generation frequency while fixing the bubble size.

Wake-induced lift acting on spherical bubbles has been theoretically and numerically investigated; however, the effective range of wake-induced lift is extremely limited. 
Moore \cite{Moore1963} proposed a wake model for a spherical bubble based on boundary layer theory, predicting that a spherical bubble has a narrow, elongated wake structure with width of $O\left({Re}^{-1/4}\right)$ and length of $O\left({Re}^{1/2}\right)$. 
Thus, the region where wake-induced lift is experienced remains highly localized. 
For bubbles rising in a chain, wake-induced lift acts on far distant trailing bubbles at the in-line position; however, any bubble with a slight lateral displacement from the in-line position experiences less lift. 
Blanco and Magnaudet \cite{Blanco1995} reported that a spherical bubble does not have a standing eddy for any Reynolds number. 
However, previous studies have not sufficiently explored whether large-scale dispersion of bubbles rising in a chain can be driven solely by wake-induced lift. 
Therefore, it is necessary to investigate the relationship between wake-induced lift and bubble dispersion while focusing solely on bubble--bubble interactions (i.e., specifically excluding effects such as bubble deformation and interfacial contamination).

Bubble--bubble interaction between spherical bubbles has been investigated numerically and analytically for several decades. 
However, most of these studies focused on the drag forces acting vertically between bubbles, such as wake-induced drag reduction. 
Harper \cite{Harper1970} analyzed the potential repulsion between two bubbles in a line. 
He predicted that a balance between this repulsion and the wake-induced attraction leads to an equilibrium distance between the bubbles. 
The potential effect decays rapidly as the separation distance between the bubbles increases. 
The existence of an equilibrium separation was numerically confirmed by Yuan and Prosperetti \cite{Yuan1994} for $20\le Re\le 200$, and analytically by Harper \cite{Harper1997} by incorporating vorticity diffusion in the wake. 
Biesheuvel and Van Wijingaarden \cite{Biesheuvel1982} investigated the relative motion of a pair of spherical bubbles interacting in the potential flow. 
Kok \cite{Kok1993} analyzed the motion of bubbles by incorporating viscous diffusion effects through the global kinetic energy balance proposed by Levich \cite{Levich1962}. 
The shear-induced lift acting on bubbles has been investigated numerically by Legendre and  Magnaudet \cite{Legendre1998} and experimentally by Tomiyama \cite{Tomiyama2002}. 
Hallez and Legendre \cite{Hallez2011} evaluated shear-induced lift acting on the trailing bubble within the wake of the leading bubble. 
This wake-induced lift laterally displaces the trailing bubble away from the centerline of the wake from the leading bubble. 
This model predicts potential and viscous interactions for a pair of bubbles at arbitrary relative positions.

Bubble--bubble interactions are related to bubble clustering.
Sangani \textit{et al.} \cite{Sangani2001} experimentally observed bubbles clustering into horizontal rafts in a vertical channel with $Re\gg1$ and $We< 1$. 
Takagi and Matsumoto \cite{Takagi2011} rationalized the formation of crescent-shaped clusters through a combined consideration of surfactant-induced coalescence suppression, lift acting on individual bubbles, and bubble--bubble interactions. 
Lee and Choi \cite{Lee2026} numerically demonstrated horizontal clustering of bubbles (equivalent diameter $d_e = 1\,\mathrm{mm}$ and $1.5\,\mathrm{mm}$) through bubble--bubble interactions. 
Many studies have investigated the relevance of bubble--bubble interaction theories to bubble cluster formation mechanisms through experiments and modeling. 
Maeda \textit{et al.} \cite{Maeda2021} reported that bubble pairs remained stable in oblique positions. 
Kusuno and Sanada \cite{Kusuno2015} demonstrated that the in-line configuration becomes unstable for observed bubble trajectories with $300\le Re\le 600$, attributing the instability to wake-induced lift. 
Overall, these studies suggest that bubble cluster formation is critically dependent on wake-induced lift in addition to attractive bubble--bubble interactions. 
However, existing studies have not sufficiently clarified either the effect of wake-induced lift in multibubble  systems (bubble chains) or the transition from in-line configurations.

\added{Research on bubble swarms, e.g., bubble plumes, typically focuses on mixing and agitation performance. 
  In the case of deformed bubbles, strong vorticity at the gas--liquid interface leads to the formation of standing eddies and vortex shedding, which notably enhance liquid mixing and agitation (Risso \cite{risso2018agitation}). 
  In contrast, spherical bubbles generate substantially weaker interfacial vorticity and do not form standing eddies. 
  As a result, their ability to promote mixing and agitation remains unclear (Moore \cite{Moore1963}, Blanco \textit{et al.} \cite{Blanco1995}). 
  Previous studies on spherical bubble-induced mixing have primarily focused on microbubbles, aiming to enhance reaction efficiency via increased gas--liquid interfacial area. 
  However, Gong \textit{et al.} \cite{gong2009effect} showed that smaller bubbles are more readily entrain into their self-induced liquid flow and tend to undergo premature discharge. 
  This behavior can counterintuitively reduce overall mass transfer efficiency. 
  Their results suggest that large-scale liquid circulation exerts dominant influence on mixing and agitation than bubble size alone. 
  Motivated by these findings, the present study examines a spherical bubble chain to elucidate the coupling mechanisms between the bubble swarm and its self-induced flow field.
  }

\replaced{In this study}{Herein}, we investigate the wake-induced dispersion for spherical bubbles rising in a chain under conditions where bubble deformation and contamination are negligible. 
Although the wake-induced lift acting on a spherical bubble has been predominantly studied theoretically, Wang and Socolofsky \cite{Wang2015} reported that the experimental behavior of bubbles in a chain changed as the flow rate increased. 
However, it remains difficult to distinguish whether these induced motions are associated with bubble size increase or the bubble generation frequency (i.e., bubble separation). 
Moreover, it is impractical to numerically simulate the entire bubble chain configuration due to the enormous computational cost. 
Resolving the boundary layer is essential for accurately predicting the lift force and lateral migration of bubbles in numerical simulations (Hallez and Legendre \cite{Hallez2011}, Zhang \textit{et al.} \cite{Zhang2017}, \cite{Zhang2019}, Kusuno and Sanada \cite{Kusuno2021}). 
Under the present conditions, bubbles rising in a chain have the Reynolds number $Re=O\left(50\right)$, bubble separation to radius ratio $S=L/R\geq20$, and field of view extending across $600R$.

This study experimentally investigates the motion of spherical bubbles with a controlled generation frequency and fixed diameter by using an originally developed bubble generation system. 
We experimentally observe the dispersion of bubbles in a chain driven by wake-induced lift under conditions where bubble deformation and contamination are negligible. 
We use a bubble chain model that includes only bubble--bubble interactions to theoretically examine the dispersion produced solely by wake-induced lift. 
We then investigate the dispersion mechanism in bubbles and evaluate the effect of lift in driving motion by combining the experimental observations with the model analyses.

\section{Experimental setup and modeling method}
\subsection{Experimental setup}

\begin{figure}%[t]
\includegraphics[width=1\linewidth]{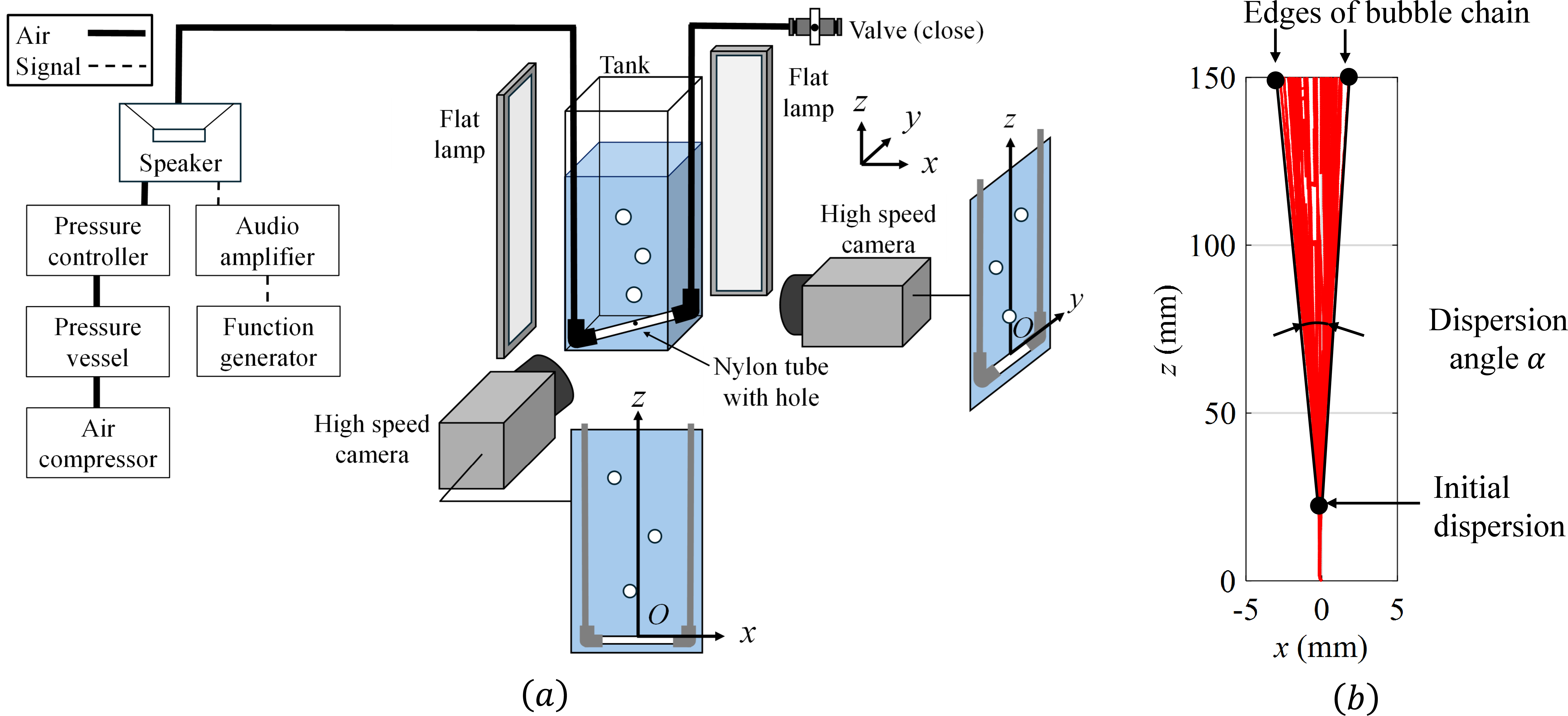}% Here is how to import EPS art
\caption{\label{fig1} (a) Schematic of the experimental setup of the bubble generation system and three-dimensional high-speed imaging. (b) Definition of dispersion angle $\alpha$ of the recorded bubble trajectories.}
\end{figure}

We investigated the dispersion of chained bubbles under negligible bubble deformation and contamination conditions. 
Figure~\ref{fig1}(a) shows a schematic of the experimental setup. 
Spherical bubbles of fixed diameter $d$ were generated at a variable bubble generation frequency $f$ using an acoustic system where bubbles emerge from an orifice (diameter $0.3\,\mathrm{mm}$) on the side of a nylon tube. 
The compressor and pressure controller pressurized the air inside the tube. 
A square wave output from the function generator was amplified before being converted into acoustic waves by the speaker. 
The acoustic wave propagated through the tube and generated bubbles at the orifice. 
Shirota \textit{et al.} \cite{Shirota2008} provide further details on the acoustic control of bubble generation. 
We used air as the gas phase and silicone oil (kinematic viscosity $\nu=1\,\mathrm{mm^2/s}$, surface tension $\sigma=16.9\,\mathrm{mN/m}$, and density $\rho_L=816\,\mathrm{kg/m^3}$) as the liquid phase to maintain clean bubble surfaces. 
The silicone oil filled the acrylic tank ($120\,\mathrm{mm} \times 120\,\mathrm{mm} \times 300\,\mathrm{mm}$). 
The orifice was placed near the center of the tank bottom. 
Takemura \textit{et al.} \cite{Takemura2002} demonstrated that wall effects are negligible when the nondimensional distance $l^\ast \approx Rel/2R > 10$, where $l$ is the distance between the bubble and the wall; $l^\ast \approx 6000$ in our experiments so the wall has a negligible effect on the bubble dynamics. The bubble diameter was fixed at $d=0.4, 0.5,$ or $0.6\,\mathrm{mm}$, while the generation frequency was set to $f=4, 8, 12,$ and $20\,\mathrm{Hz}$. 

We used two complementary imaging methods to observe the bubbles. 
First, two high-speed cameras captured the three-dimensional bubble motion from orthogonal directions. 
Each camera provided a resolution of 2560 × 2048 pixels (0.058 mm/pixel) at 250 fps. 
We defined the bubble generation point as the origin and set the vertical direction as the $z-$axis. 
The respective camera projection planes corresponded to the $x-z$ and $y-z$ planes. 
Second, a single high-speed camera with a higher (1080 fps) frame rate captured the detailed dynamics of bubbles with different diameters. This camera captured the projection of bubbles onto the $x-z$ plane at a resolution of 2048 × 2048 pixels (0.078 mm/pixel). 
We recorded images after bubbles reached a steady state. 

The Reynolds number $Re$ and the normalized bubble separation $S$ characterizing the bubbles rising in a chain are defined as 

\begin{equation}
Re=\frac{Ud}{\nu}
\label{Re_definition},
\end{equation}
\begin{equation}
S=\frac{2U}{fd}
\label{S_definition},
\end{equation}
where $U$, $d$, $\nu$, and $f$ are the vertical rising velocity, averaged equivalent diameter, kinematic viscosity of the liquid, and bubble generation frequency, respectively. 
The Reynolds numbers in the experiment were within the range $28\le Re\le 52$ for $d=0.5\,\textrm{mm}$. 
Note that the Reynolds number $Re$ varies due to the dispersion-induced variation in rising velocity (Fig.~\ref{fig8}) and the frequency dependence of the upward flow, even for the fixed bubble diameter $d$. 
\added{The Bond number, $Bo\approx 0.1$, is based on the bubble diameter, and the Weber number, which is calculated using the bubble diameter and the terminal rise velocity, falls within $0.1\le We\le 0.3$. 
  According to the bubble shape regime map proposed by Clift \textit{et al.} \cite{clift1978bubbles}, bubbles under these conditions can be classified as spherical, corresponding to a Reynolds number range of $28\le Re\le 52$.
  }
We use a dispersion angle $\alpha$ introduced in Ref.~\cite{Atasi2023} to quantify the dispersion in the $x-z$ plane, as shown in Fig.~\ref{fig1}(b). 
The dispersion angle corresponds to the subtended angle between the initial point for bubble dispersion and the two edges at the top of the field of view ($z=150\,\mathrm{mm}$). 
We defined the initial dispersion point as the lowest $z-$position at which the width of the bubble chain exceeds $0.1\,\mathrm{mm}$.

\subsection{Modeling method}
\subsubsection{Calculation method of the bubble chain model}

\begin{figure}%[t]
\includegraphics[width=0.8\linewidth]{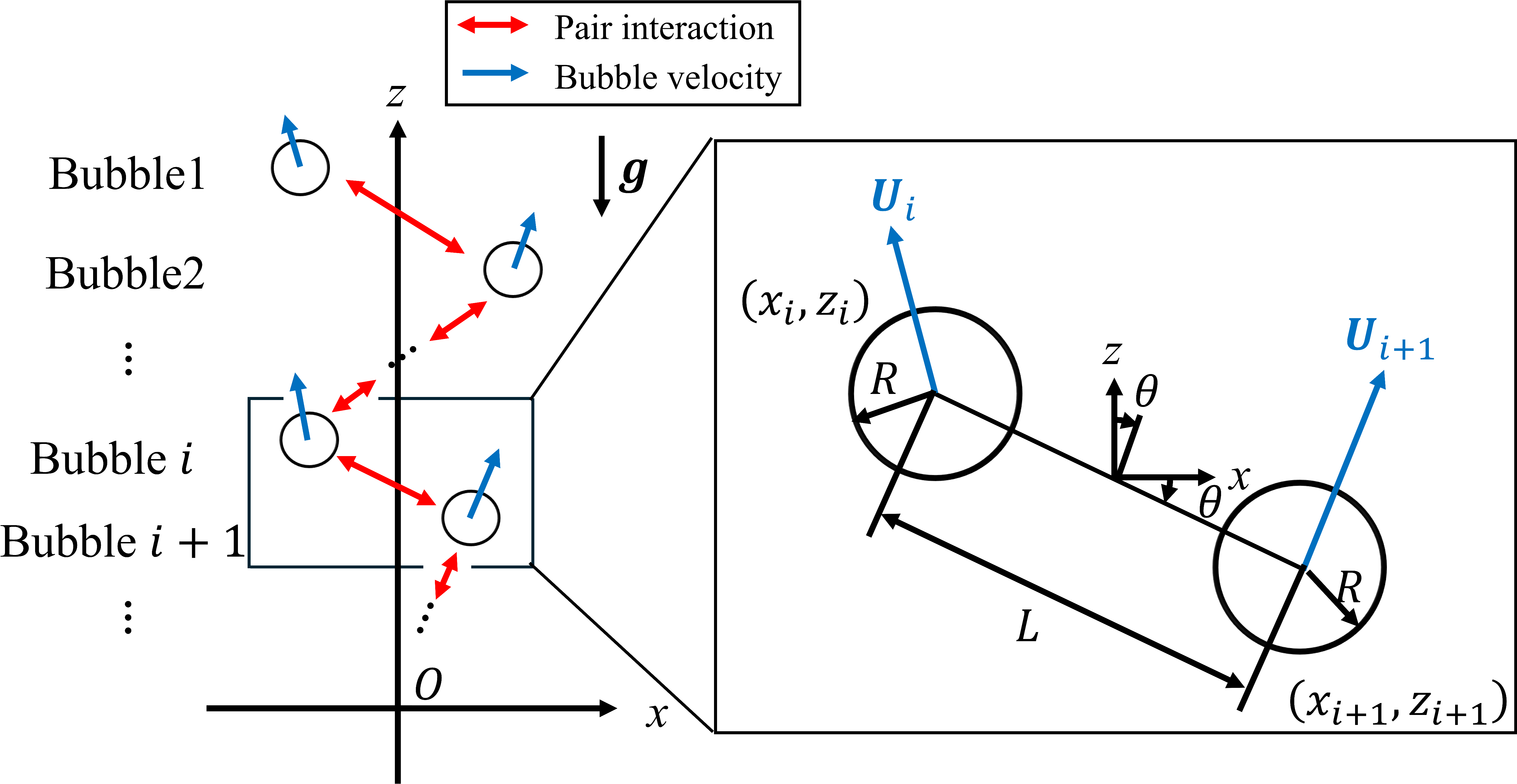}% Here is how to import EPS art
\caption{\label{fig2} Schematic of the bubble chain model incorporating interactions with adjacent bubbles.}
\end{figure}

We developed a model for clean spherical bubbles rising in a chain to evaluate the dispersion induced by bubble--bubble interactions (Fig.~\ref{fig2}). 
Each bubble in a chain disperses through interactions with adjacent bubbles, as formulated by Hallez and Legendre \cite{Hallez2011}.  
The model tracks trajectories in a Lagrangian framework by solving the equations of motion for each bubble, including the interaction forces, to predict the dispersion. 
We labeled the bubbles in order of generation. 
The equation of motion for bubble $i$ is:

\begin{equation}
\frac{4}{3}\pi \rho_G R^3 \frac{d\bm{U}_i}{dt} = \bm{F}_B + \bm{F}_{QS,i} + \bm{F}_{AM,i} + \bm{F}_{I,i,i-1} + \bm{F}_{I,i,i+1},
\label{eq3}
\end{equation}
where $R$, $\bm{U}_i$, $\bm{F}_B$, $\bm{F}_{QS,i}$, $\bm{F}_{AM,i}$, and $\bm{F}_{I,i,j}$  $(j=i-1,i+1)$ are the bubble radius, velocity, buoyancy force, steady drag force acting on an isolated bubble, added-mass force, and interaction force exerted on bubble $i$ by bubble $j$, respectively. 
Maeda \textit{et al.} \cite{Maeda2021} and Kusuno and Sanada \cite{Kusuno2015} provide further details on the formulations of the interaction between pairwise bubbles. 
Our model considers only the interaction forces induced by the adjacent bubbles $i-1$, $i+1$, neglecting interactions with more distant bubbles because of their small magnitudes. 
The interaction force term is formulated as:

\begin{equation}
\bm{F}_{I,i,j} = \frac{1}{2} \rho_L \pi R^2 |\bm{U}_i \cdot \bm{e}_z|^2 (-C_{DI,i,j} \bm{e}_z + C_{LI,i,j} \bm{e}_x),
\label{eq4}
\end{equation}
where $C_{DI,i,j}$ and $C_{LI,i,j}$ are the respective drag and lift coefficients induced by bubble--bubble interactions. 
These coefficients account for both potential effects and viscous interactions. 
We adopted the coefficient values proposed by Hallez and Legendre \cite{Hallez2011}. 
Interaction force coefficients are calculated by the average Reynolds number $Re_{pair} = R(|\bm{U}_i| + |\bm{U}_{i+1}|) / 2\nu$, the dimensionless separating distance $S=L/R$, and the relative angle $\theta$ for bubble pair ($i$, $i+1$).

We tracked 50 bubbles in the model, starting with bubble 1 generated at $t=0\,\mathrm{s}$. 
Subsequent bubbles were generated at intervals of $1/f\,\mathrm{s}$ such that bubble $i$ was generated at $t=(i-1)/f\,\mathrm{s}$. 
Only the first, second, third, and fourth terms in Eq.~(\ref{eq3}) act on a newly generated bubble $i$, until bubble $i+1$ is generated at $t=i/f\,\mathrm{s}$ and the final term acts on bubble $i$.

\replaced{
  To initiate dispersion within the bubble chain, we randomly assigned the initial lateral bubble positions. 
  Specifically, the initial $x-$coordinate of each bubble was randomly assigned within the range $-0.005\le x_0\le 0.005\,\mathrm{mm}$ prior to computation. 
  As demonstrated by Hallez and Legendre \cite{Hallez2011}, perfectly aligned bubbles experience no lift force; thus, this imposed randomness generates the wake-induced lift necessary to trigger bubble interactions. 
  As discussed in the Appendix, these perturbations serve only to initiate interactions and do not influence the subsequent dispersion behavior. 
  The initial vertical position was set to $z_0 = 0 \,\mathrm{mm}$, and the initial rise velocity to $U_0 = 10 \,\mathrm{mm/s}$.
  }
  {
  We specified initial conditions to solve the differential equation in Eq.~(\ref{eq3}). 
    At generation, each bubble had an initial rising velocity of $U_0=10 \,\mathrm{mm/s}$ in the $z-$direction. 
  We randomly assigned the initial position of each bubble within the range $-0.005\le x_0\le 0.005\,\mathrm{mm}$ and at a fixed $z_0=0\,\mathrm{mm}$. 
  This randomness slightly displaced the relative bubble positions from a perfectly in-line configuration, inducing lateral lift forces through bubble--bubble interactions.
  }

\subsubsection{Incorporating the upward flow effect into the model}

\begin{figure}%[t]
\includegraphics[width=0.8\linewidth]{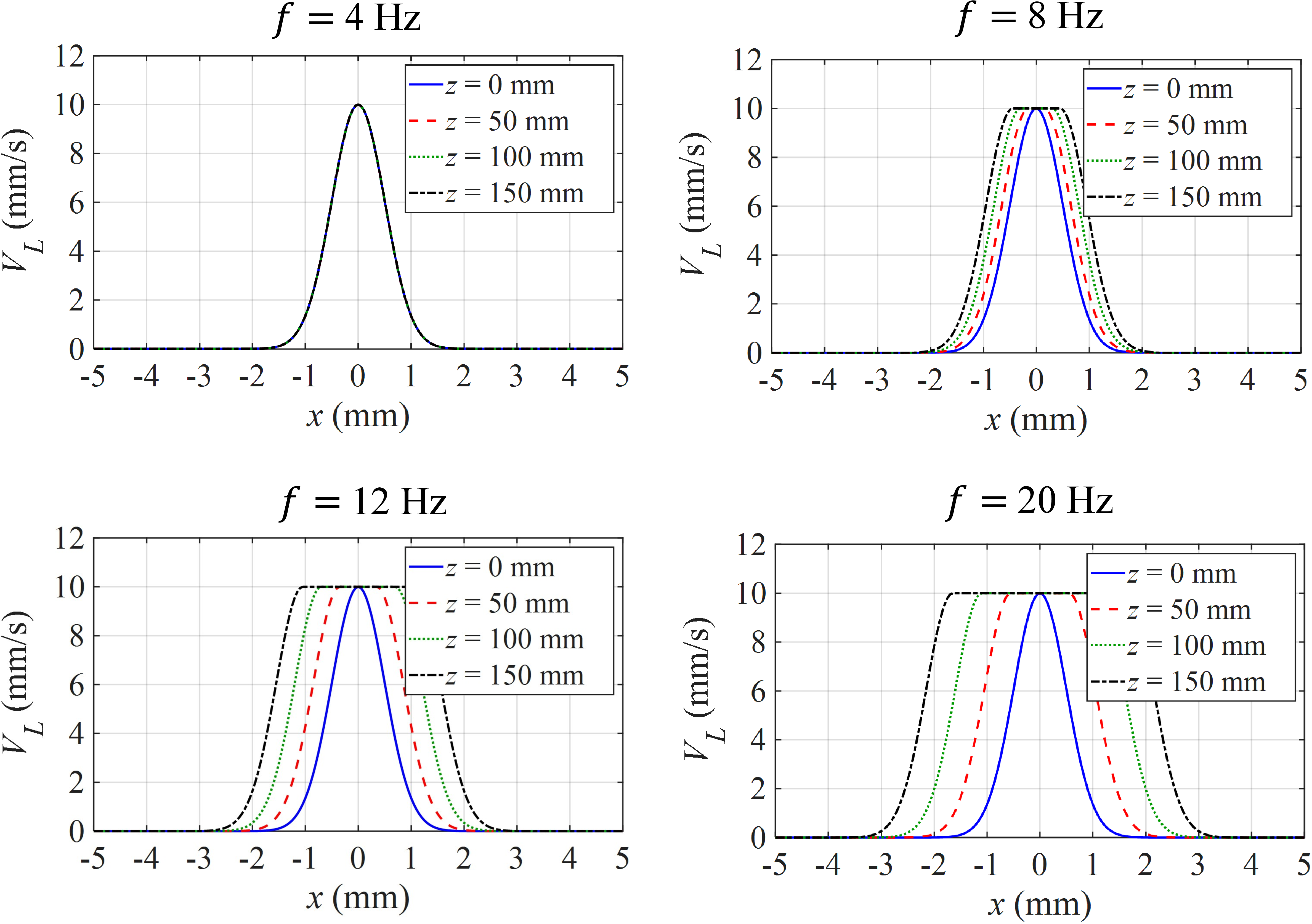}% Here is how to import EPS art
\caption{\label{fig3} Assumed upward flow profiles at different heights. The edge regions are modeled by a Gaussian distribution with a standard deviation $\sigma=0.5\,\mathrm{mm}$, while the central flat region expands with height $z$.}
\end{figure}

We examined the effect of upward flow on the bubble dispersion. 
Incorporating the effect of the upward flow into the model demonstrated that the second stage of dispersion is induced by the upward flow. 
The shear-induced lift acting on a single spherical bubble can be expressed as proposed by Auton \cite{Auton1987} and Legendre and Magnaudet \cite{Legendre1998}:

\begin{equation}
\bm{F}_{LF} = C_{LF} \frac{4\pi R^3}{3} \rho_L (\bm{V}_L - \bm{U}) \times (\nabla \times \bm{V}_L),
\label{eq5}
\end{equation}
where $\bm{U}$ is the bubble velocity, $\bm{V}_L$ is the velocity field evaluated at the center of the bubble, and $C_{LF}$ is the lift coefficient induced by shear flow. 
Legendre and Magnaudet \cite{Legendre1998} expressed the lift coefficient as $C_{LF}^{LM} = C_L^A (Re + 16) / (Re + 29)$ for viscous fluids. 
Assuming that the upward flow has only a vertical component in the two-dimensional model, the transverse lift acting on a bubble is:

\begin{equation}
F_{LFx} = C_{LF} \frac{4\pi R^3}{3} \rho_L (V_L - U) \frac{\partial V_L}{\partial x},
\label{eq6}
\end{equation}
where $V_L$ is the vertical flow velocity and $U$ is the rising velocity. 
The upward flow velocity profile is prescribed within the computational domain of the model. 
For each time step, the lift force acting on each bubble is calculated from Eq.~(\ref{eq6}) at its instantaneous position. 
The results are then substituted into the equations of motion (Eq.~(\ref{eq3})), thereby incorporating the upward flow effect into the model as an additional term $\bm{F}_{LF}=(F_{LFx}, 0)$. For bubble $i$, this gives:

\begin{equation}
\frac{4}{3}\pi \rho_G R^3 \frac{d\bm{U}_i}{dt} = \bm{F}_B + \bm{F}_{QS,i} + \bm{F}_{AM,i} + \bm{F}_{I,i,i-1} + \bm{F}_{I,i,i+1} + \bm{F}_{LF},
\label{eq7}
\end{equation}
\added{Figure~\ref{fig3} shows the upward flow profile for each frequency $f$. We prescribed the velocity distribution of the upward flow to replicate the experimental dispersion angle $\alpha$ within the model. 
  This approach assumes that the maximum upward flow velocity remains sufficiently small compared with the bubble rise velocity, and that the upward flow spreads laterally as the bubble chain disperses during ascent, as detailed in the Appendix.
}
The maximum velocity is assumed to be sufficiently smaller than the bubble terminal velocity (100 mm/s, see Fig.~\ref{fig8}(b)) and is set to $10\%$ of it, i.e., 10 mm/s. 

The shear region of the upward flow shifts outward to both sides as $z$ increases. 
The shifted shear region means that dispersion increases since bubbles experience lift force in the shear region. 
We define the upward flow profile as:

\begin{equation}
V_L(x, z) = 
\begin{cases} 
V_{L0} & |x| \le w(z) \\
V_{L0} \exp \left( -\frac{(|x| - w(z))^2}{2\sigma^2} \right) & |x| \ge w(z)
\end{cases},
\label{eq8}
\end{equation}
where $V_{L0}=10 \,\mathrm{mm/s}$ is the maximum upward flow velocity, $\sigma=0.5\,\mathrm{mm}$ is the standard deviation  of the Gaussian component, and $w(z)=\beta z$ is the half-width of the maximal velocity region at height $z$. 
The expansion rate $\beta$ is determined for each frequency such that it reproduces the experimentally obtained dispersion angles $\alpha$ for the 0.5 mm bubbles (listed in Table~\ref{tab:UpwardflowParameter}).

\begin{table}[b]
\caption{\label{tab:UpwardflowParameter}%
Flat width of upward flow depends on the generation frequency $f$.
}
\begin{ruledtabular}
\begin{tabular}{cc}
\textrm{$f\,\mathrm{[Hz]}$}&
\textrm{$\beta\,\mathrm{[-]}$}\\
\colrule
4 & 0\\
8 & 0.003\\
12 & 0.007\\
20 & 0.011\\
\end{tabular}
\end{ruledtabular}
\end{table}

\section{Results}
\subsection{Experimental results}
\subsubsection{Three-dimensional dispersion of bubbles rising in a chain}

\begin{figure}%[t]
\includegraphics[width=1\linewidth]{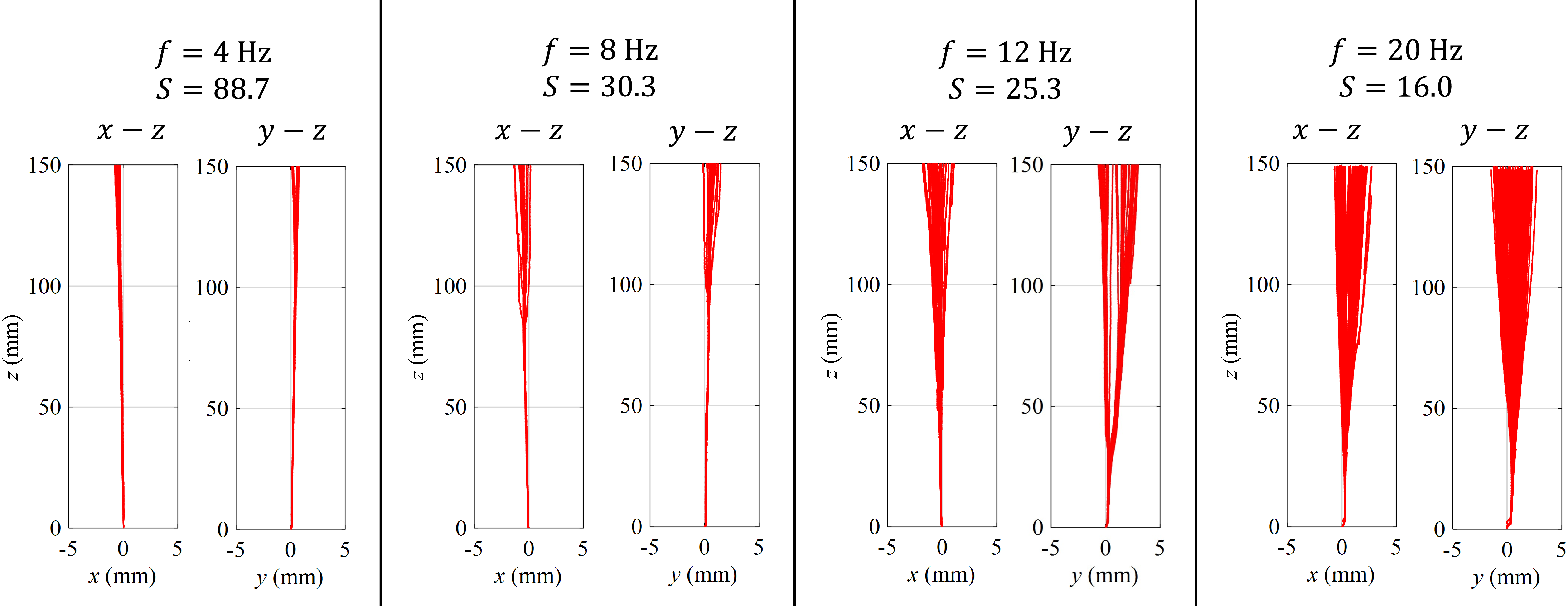}% Here is how to import EPS art
\caption{\label{fig4} Trajectories of bubbles rising in a chain ($d=0.5\,\mathrm{mm}, \nu=1 \,\mathrm{mm^2/s}$).}
\end{figure}

Figure~\ref{fig4} shows the trajectories of bubbles rising in a chain in silicone oil ($\nu=1\,\mathrm{mm^2/s}$), with a fixed bubble diameter $d=0.5\,\mathrm{mm}$.
We captured the bubble motion from directions perpendicular to the $x-z$ and $y-z$ planes.  
\added{The plot includes 26, 60, 77, and 132 bubbles for the frequencies of $f= 4, 8, 12,$ and $20\,\mathrm{Hz}$, respectively. 
  Each dataset represents the plotted results from a single experimental trial.
  }
At $f=4\,\mathrm{Hz}$, each bubble ascends almost vertically, forming a stable chain. On the other hand, for $f\ge8\,\mathrm{Hz}$, the trajectories deviate from the vertical axis. 
The rising bubbles developed a large-scale (relative to the bubble diameter) V-shaped dispersion structure, even under conditions in which a spherical bubble shape was maintained.
\added{As presented in the Appendix, the trajectory asymmetry of the bubble chain originates from randomness during bubble generation.
}

We controlled the bubble separation through modulating the generation frequency $f$. 
The dimensionless bubble separation $S$ calculated from Eq.~\ref{S_definition} for each frequency is given in Fig.~\ref{fig4}.
The wake length is $O\left({Re}^{1/2}\right)\sim7$ based on the scaling proposed by Moore \cite{Moore1963}, which is less than $S$ for all frequencies. 
Therefore, no significant wake-induced lift is expected. Despite this, a large-scale dispersion structure was observed with a clear frequency dependence.

\begin{figure}%[t]
\centering
\includegraphics[width=0.6\linewidth]{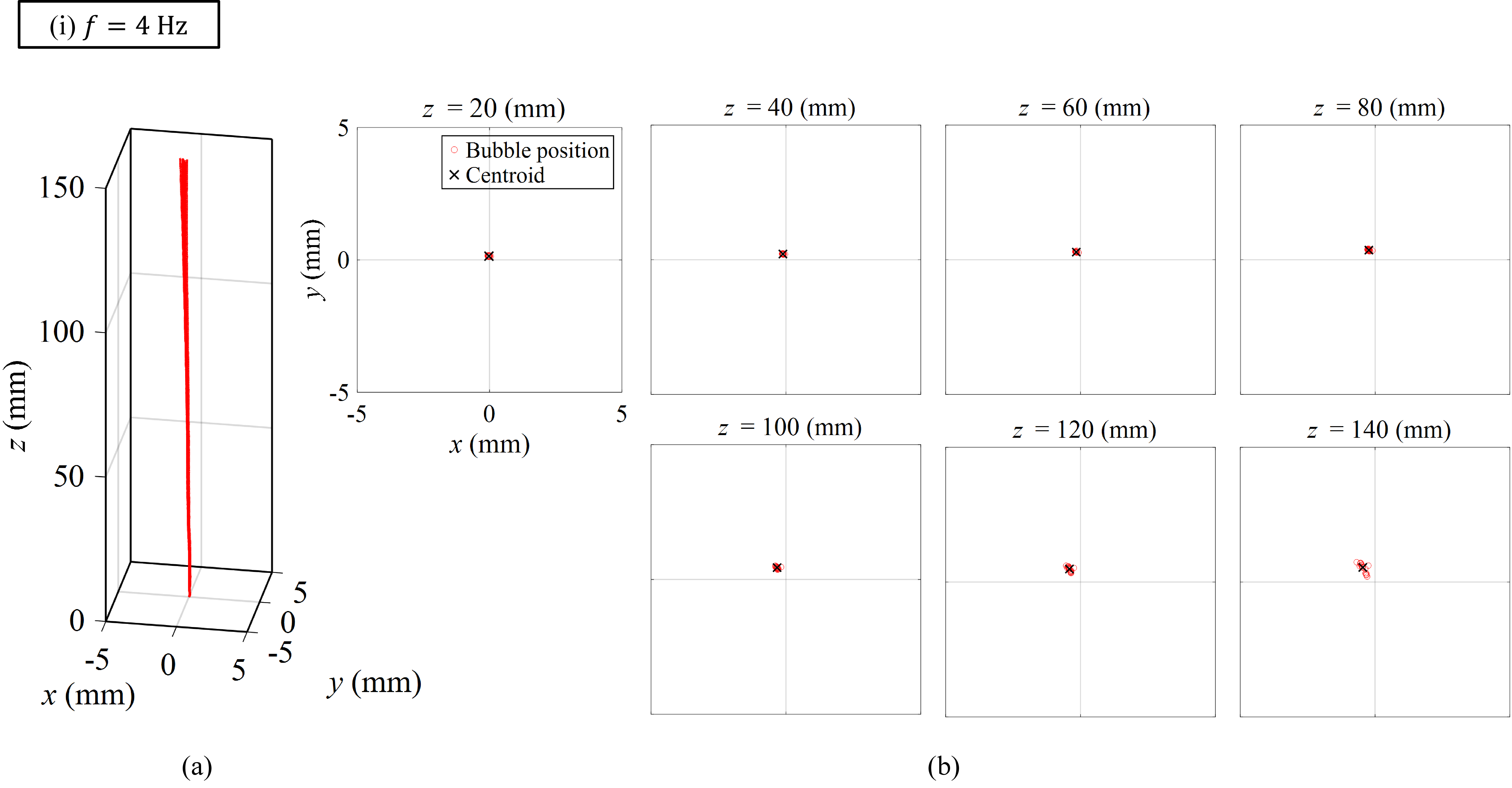}\\% Here is how to import EPS art
\includegraphics[width=0.6\linewidth]{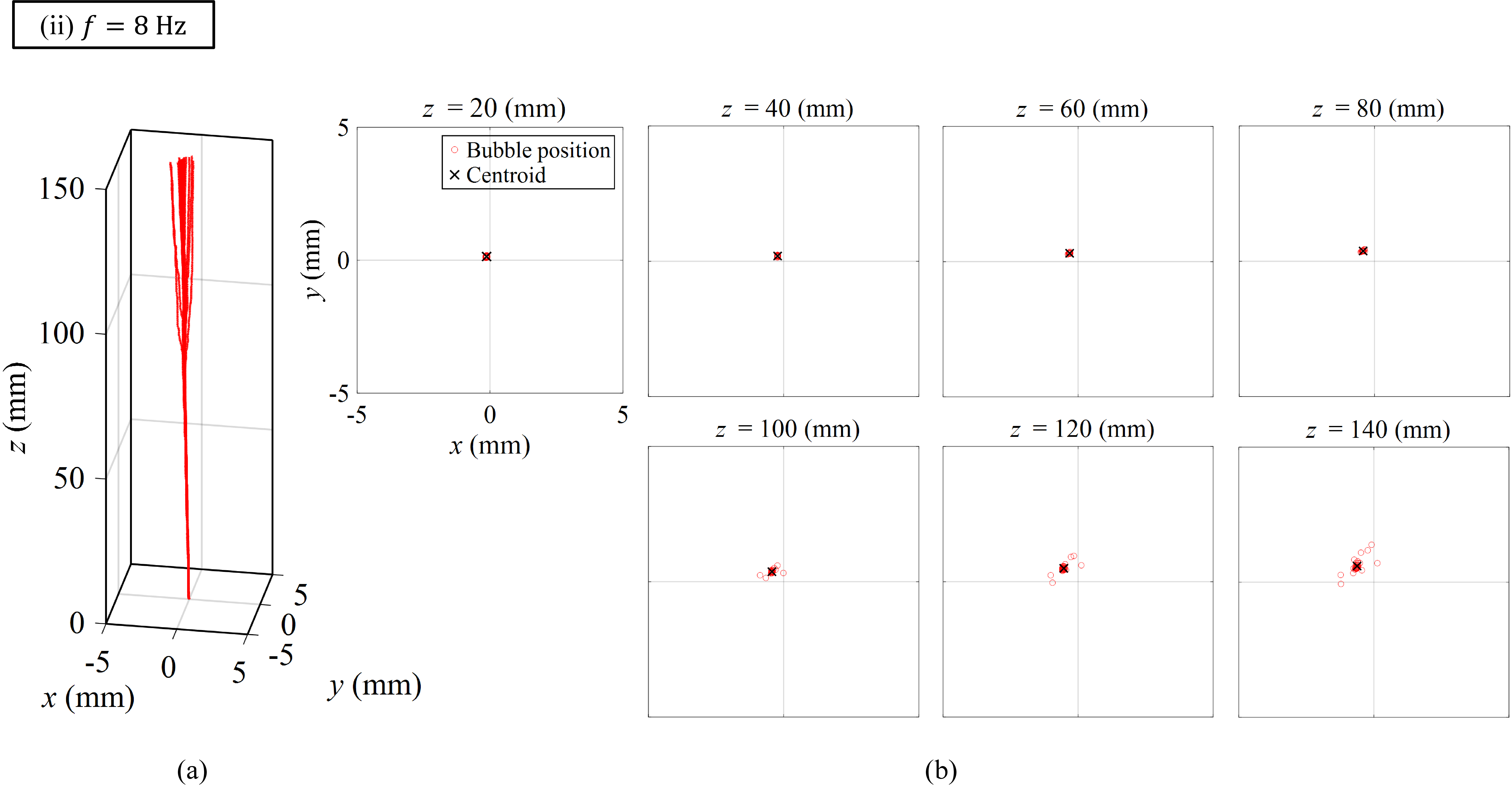}\\% Here is how to import EPS art
\includegraphics[width=0.6\linewidth]{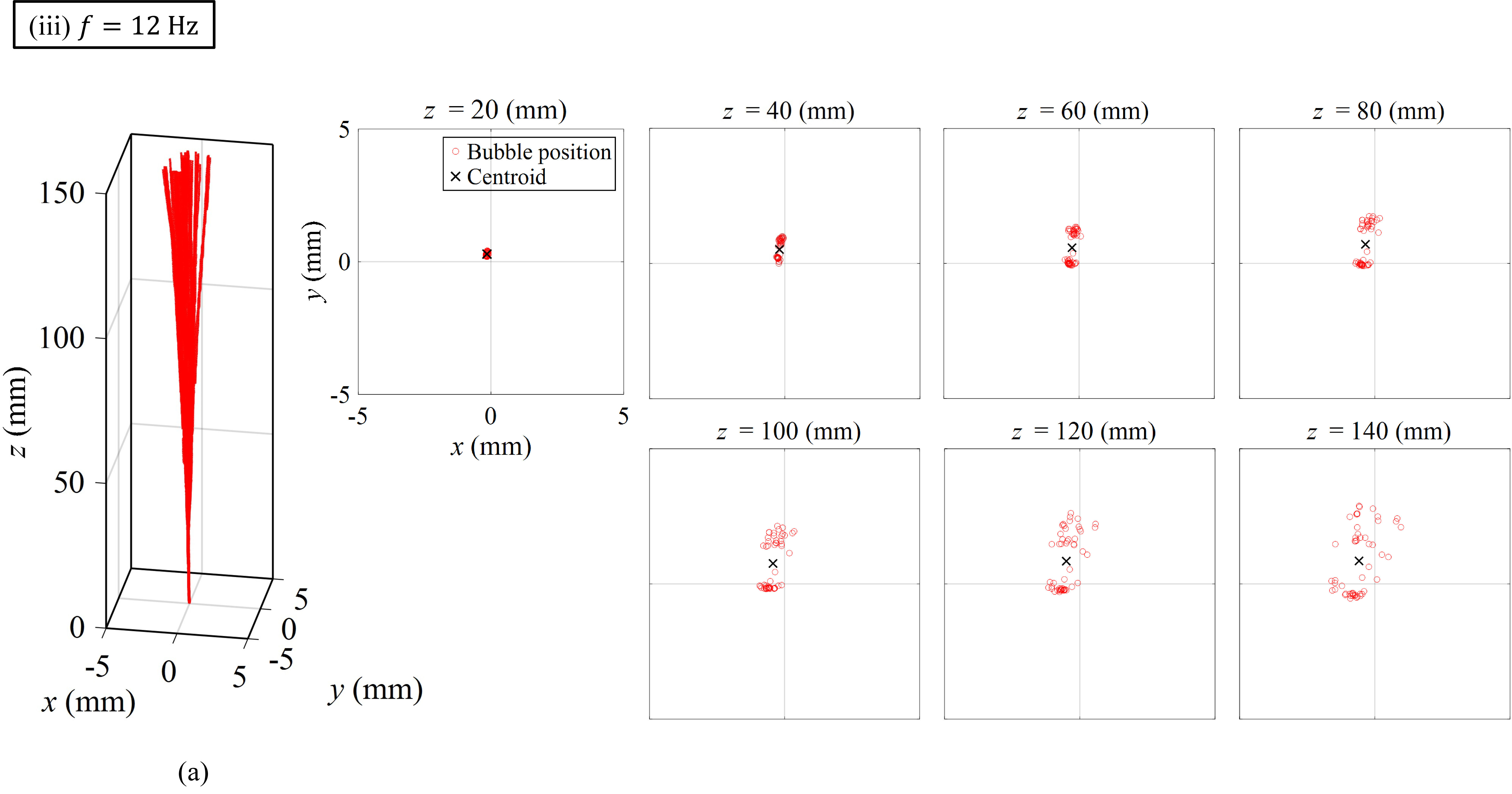}\\% Here is how to import EPS art
\includegraphics[width=0.6\linewidth]{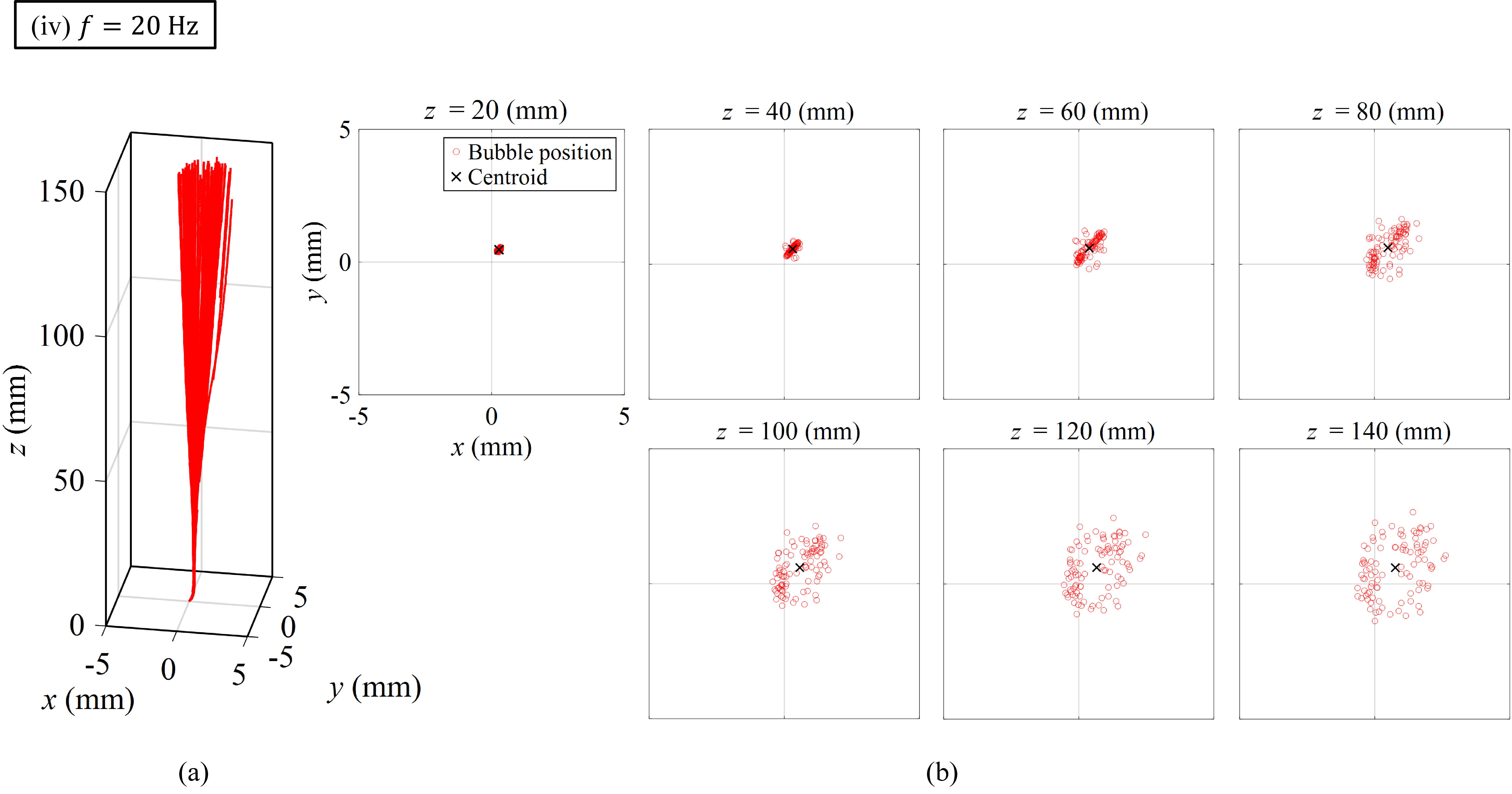}% Here is how to import EPS art
\caption{\label{fig5} (a) Three-dimensional dispersions and (b) the corresponding bubble passage distributions in cross-sections at 20 mm intervals for bubble generation frequencies of: (i) $f=4\,\mathrm{Hz}$, (ii) $f=8\,\mathrm{Hz}$, (iii) $f=12\,\mathrm{Hz}$, and (iv) $f=20\,\mathrm{Hz}$. The centroid of the distributions in (b) is marked with a cross.}
\end{figure}

Figure~\ref{fig5}(a) shows the three-dimensional trajectories of individual bubbles and Fig.~\ref{fig5}(b) shows the bubble passage positions on horizontal cross-sections at 20 mm intervals. 
Although the bubbles are relatively stable and the overall dispersion is small at $f=4\,\mathrm{Hz}$, the dispersion exhibits anisotropy. 
The bubbles remain stable up to $z=100\,\mathrm{mm}$, but some bubbles deviate from the centroid of the distribution for $z>100\,\mathrm{mm}$, and the dispersion range slightly expands along the approximately $120^{\circ}$ direction. 
At $f=8\,\mathrm{Hz}$, bubbles show slightly greater dispersion while otherwise maintaining the same trends observed at $f=4\,\mathrm{Hz}$. 
The dispersion region expanding along the $45^{\circ}$ direction for $z>80\,\mathrm{mm}$. 
Nevertheless, many bubbles remain near the centroid and maintain a partially stable chain. 
At $f=12\,\mathrm{Hz}$, bubbles exhibited a two-stage dispersion. 
A stable chain was formed up to $z=20\,\mathrm{mm}$. 
An initial dispersion stage then proceeds along the $90^{\circ}$ direction, with the bubble passage positions concentrated at two separate streams by $z=80\,\mathrm{mm}$. 
\added{The individual bubbles separated into two streams in a generally alternating manner.
}
The second dispersion stage separately expands the two concentrated regions for $z>80\,\mathrm{mm}$, such that the dispersion region forms an ellipse elongated along the $90^{\circ}$ direction. 
Few bubbles remain near the centroid, while many bubbles accumulate at the edge of the dispersion region. 
At $f=20\,\mathrm{Hz}$, greater dispersion is seen while maintaining the two-stage dispersion. 
The chain remains stable up to $z=20\,\mathrm{mm}$, before the dispersion region expands along the $45^{\circ}$ direction, concentrating at two separate streams by $z=80\,\mathrm{mm}$ \added{in a generally alternating manner}, and then gradually dispersing further for $z>80\,\mathrm{mm}$. 
The bubble passages concentrate near the edge of an elliptical dispersion region with few bubbles remaining near the centroid.

\subsubsection{Frequency and bubble diameter dependence}

\begin{figure}%[t]
\includegraphics[width=0.6\linewidth]{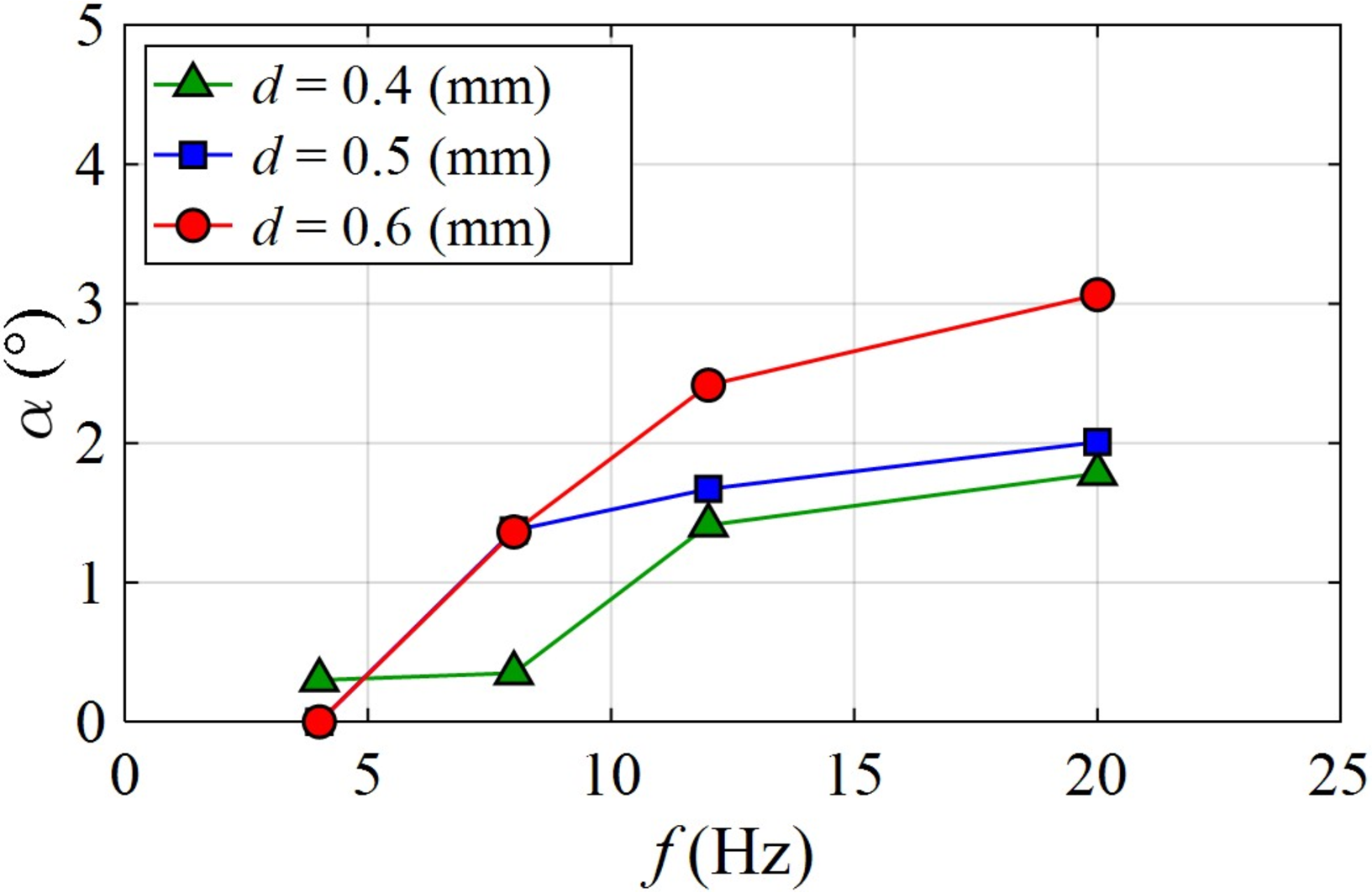}% Here is how to import EPS art
\caption{\label{fig6} Relationship between dispersion angle $\alpha$ and generation frequency $f$ for different bubble diameters.}
\end{figure}

The bubble diameter $d$ for bubbles maintaining a spherical shape was varied and the effects on the dispersion behavior were observed. 
Figure~\ref{fig6} shows the relationship between dispersion and frequency $f$ for diameters $d= 0.4, 0.5,$ and $0.6\,\mathrm{mm}$, using the dispersion angle $\alpha$ as a parameter to quantify the dispersion. 
The dispersion exhibited frequency dependence for all bubble diameters, with larger-scale dispersion structures formed at higher frequencies. 
For $f\ge 12\,\mathrm{Hz}$, the dispersion increased with the bubble diameter $d$. 
The Reynolds number $Re$ increases as the bubble diameter $d$ increases, leading to an extension of the wake length $O\left({Re}^{1/2}\right)$ and a corresponding increase to the dispersion \cite{Moore1963}. 
These results suggest that the formation of large-scale dispersion structures and the frequency dependence of the dispersion are universal phenomena observed in clean spherical bubbles rising in a chain at intermediate Reynolds number.
\added{The frequency dependence qualitatively agrees with the clean condition reported by Atasi \textit{et al.} \cite{Atasi2023}. 
  However, the present study yields smaller values for the absolute dispersion angle and its slope. 
  We attribute this discrepancy to the spherical shape constraint imposed on the bubbles in this study, which reduces the bubble size, thereby reducing the lift coefficient and yielding a smaller dispersion angle $\alpha$ with a lower rate of increase.
  }

\subsection{Model prediction}
\subsubsection{Prediction of dispersion induced by bubble--bubble interactions}

\begin{figure}%[t]
\includegraphics[width=0.7\linewidth]{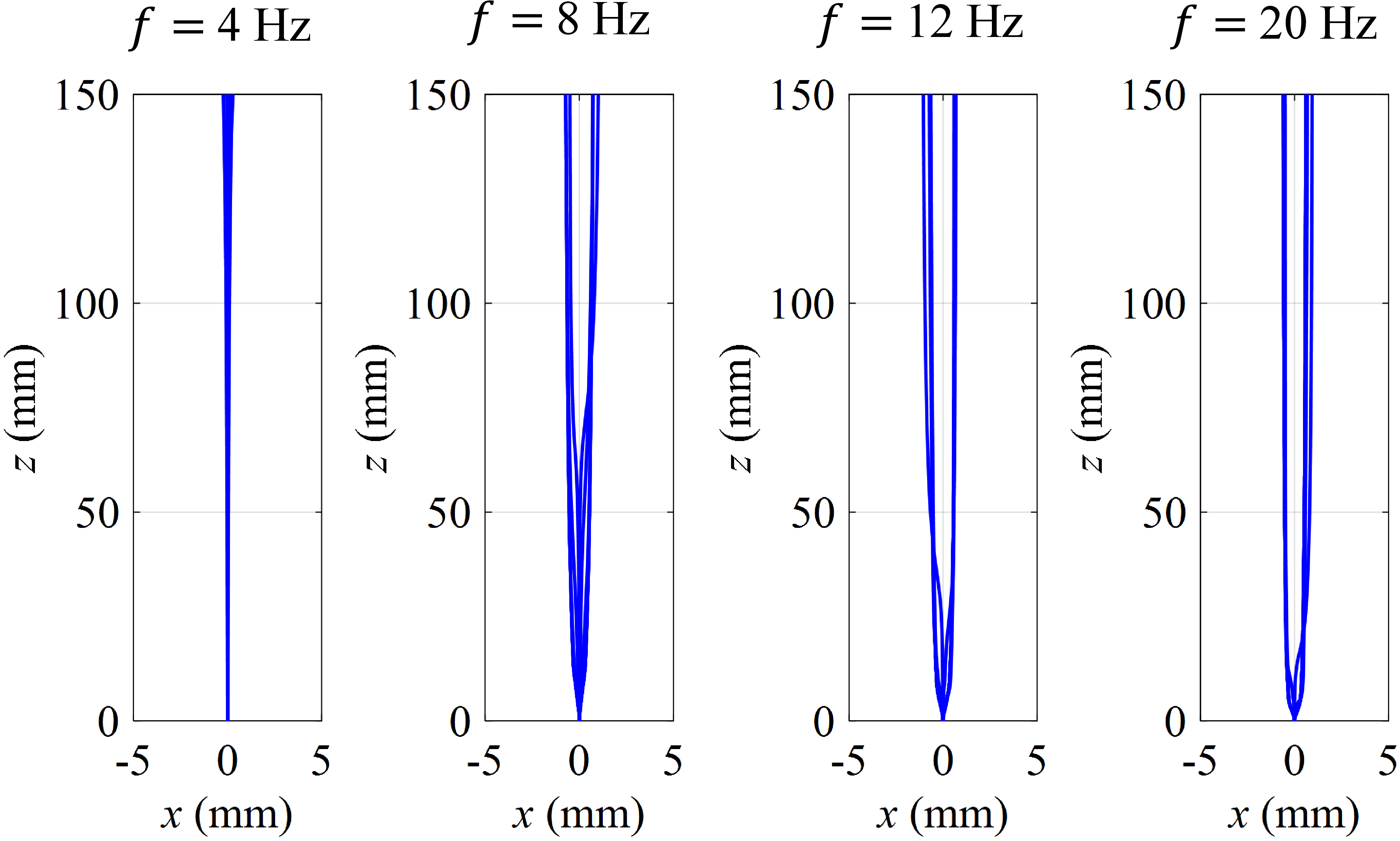}% Here is how to import EPS art
\caption{\label{fig7} Trajectories of fixed diameter bubbles ($d=0.5\,\mathrm{mm}$) in the model. The generation frequency $f$ increases from left to right.}
\end{figure}

Figure~\ref{fig7} shows the trajectories through silicone oil ($\nu=1\,\mathrm{mm^2/s}$) of bubbles ($d=0.5\,\mathrm{mm}$) in the model at $f= 4, 8, 12,$ and $20\,\mathrm{Hz}$. 
The model only accounts for bubble--bubble interactions. 
At $f=4\,\mathrm{Hz}$, each bubble rises linearly following the same approximate path, with only slight dispersion near the upper region. 
\replaced{
  However, for $f\ge 8\,\mathrm{Hz}$, each bubble alternatively moves to the opposite direction into two streams and subsequently rise vertically, forming a small-scale U-shaped dispersion.
}
{
  For $f\ge 8\,\mathrm{Hz}$, however, the bubbles first diverge laterally into two streams and subsequently rise vertically, forming a small-scale U-shaped dispersion. 
}
For $f\ge 8\,\mathrm{Hz}$, the model does not predict a significant change in dispersion width with frequency. 

\begin{figure}%[t]
\includegraphics[width=1\linewidth]{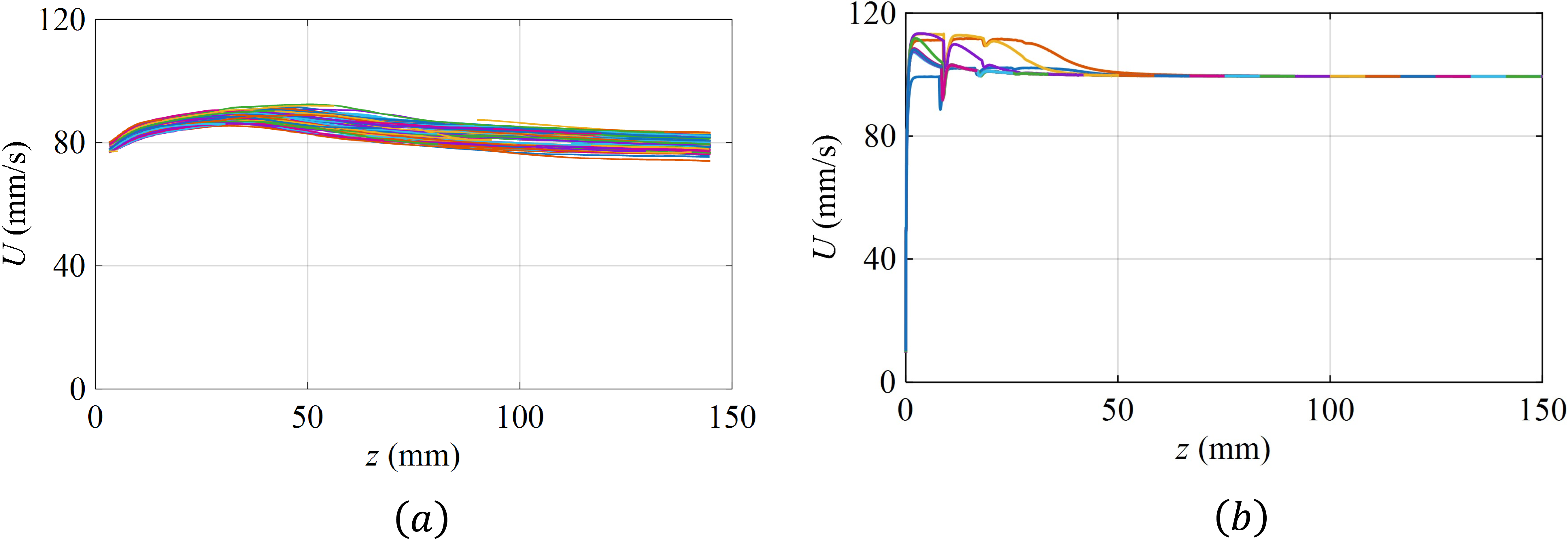}% Here is how to import EPS art
\caption{\label{fig8} Relationship between bubble rising velocity and height obtained from: (a) experiment, and (b) modeling. ($f=20\,\mathrm{Hz}$, $d=0.5\,\mathrm{mm}$, Silicone oil ($\nu=1 \,\mathrm{mm^2/s}$)).}
\end{figure}

We validated the model by comparing the bubble rising velocity with experimental measurements. 
Figure~\ref{fig8} shows the experimental and modeled evolutions of the rising velocities $U$ with height $z$.
In the experiments, each bubble accelerated from its initial velocity, after which it underwent gradual deceleration. 
We attribute the initial acceleration to drag reduction caused by reduced pressure in the wake from the leading bubble. 
Subsequently, drag increases as the bubbles disperse beyond the wake region, leading to the gradual deceleration toward a terminal velocity of approximately $U\approx 80 \,\mathrm{mm/s}$. 
The model reproduces similar acceleration and deceleration phases induced by the leading bubble wake and predicts a terminal velocity of approximately $U\approx 100 \,\mathrm{mm/s}$. 
We regard the model as valid for analyzing bubble--bubble interaction effects as it captures the qualitative trends observed experimentally.

\subsubsection{Prediction of dispersion induced by upward flow}

\begin{figure}%[t]
\includegraphics[width=0.7\linewidth]{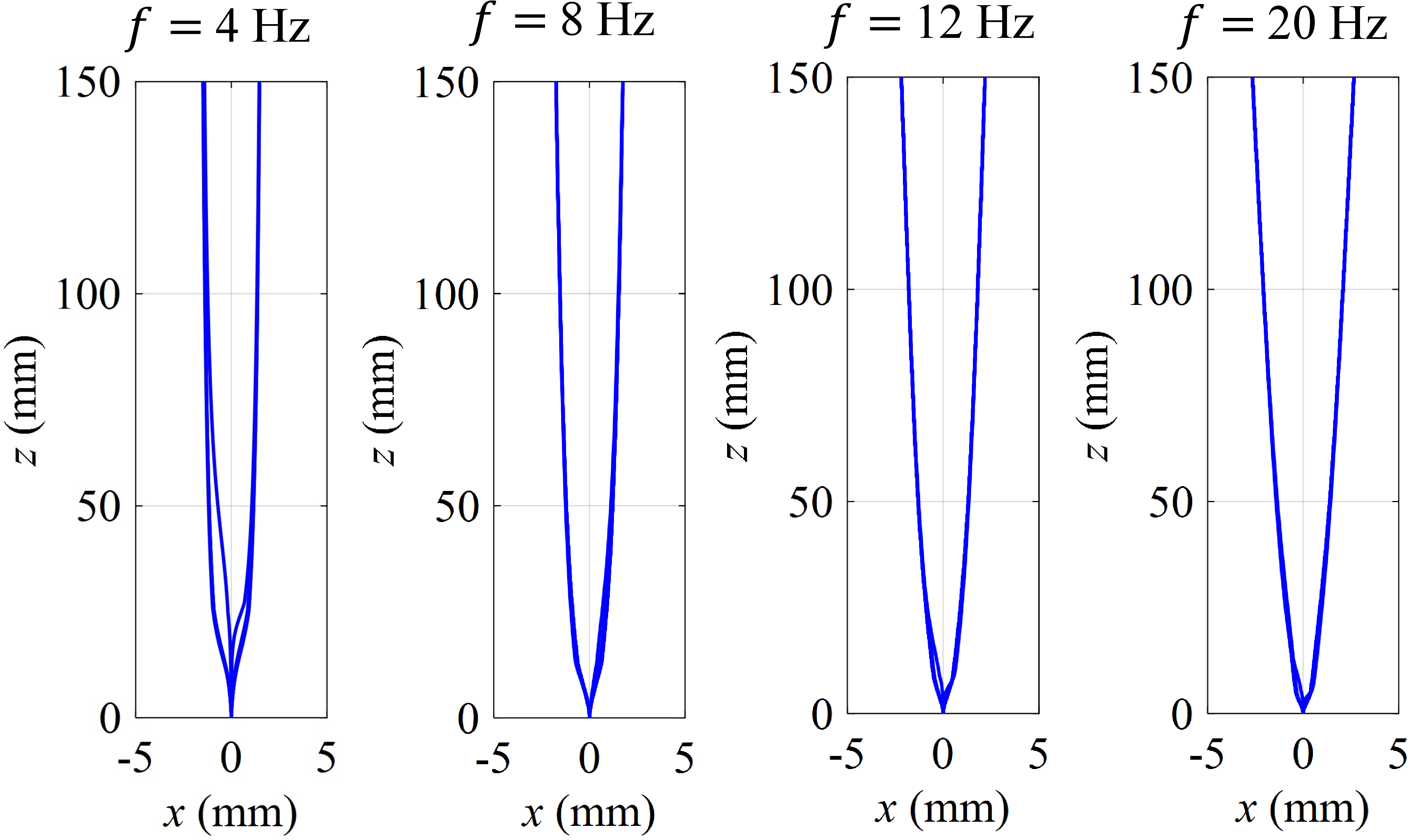}% Here is how to import EPS art
\caption{\label{fig9} Trajectories of $0.5\,\mathrm{mm}$ diameter bubbles in the extended model including upward flow effects. The generation frequency $f$ increases from left to right.}
\end{figure}

Figure~\ref{fig9} shows the modeled trajectories of bubbles ($d=0.5\,\mathrm{mm}$) generated at $f= 4, 8, 12,$ and $20\,\mathrm{Hz}$, from the extended model incorporating both bubble--bubble interaction and upward flow effects.
The overall dispersion increased when incorporating the upward flow, resulting in the formation of a large-scale V-shaped dispersion structure. 
Also, the dispersion width increased with increasing $f$ for $f\ge 8\,\mathrm{Hz}$. 

\section{Discussion}
\subsection{Dispersion induced by bubble--bubble interactions and upward flow}

\begin{figure}%[t]
\includegraphics[width=0.7\linewidth]{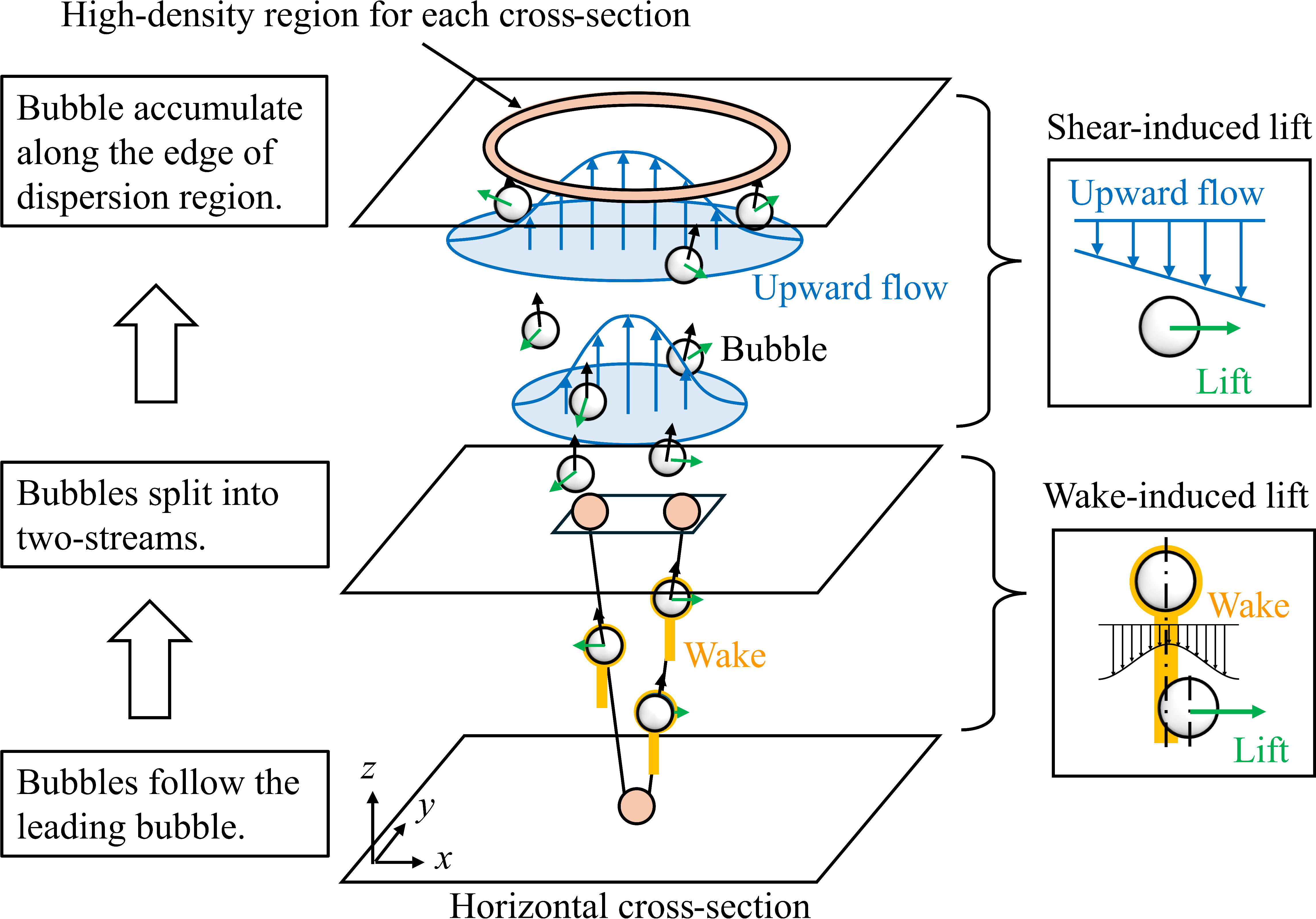}% Here is how to import EPS art
\caption{\label{fig10} \added{Schematic of the three-dimensional dispersion of bubbles rising in a chain.}}
\end{figure}

\begin{figure}%[t]
\includegraphics[width=0.95\linewidth]{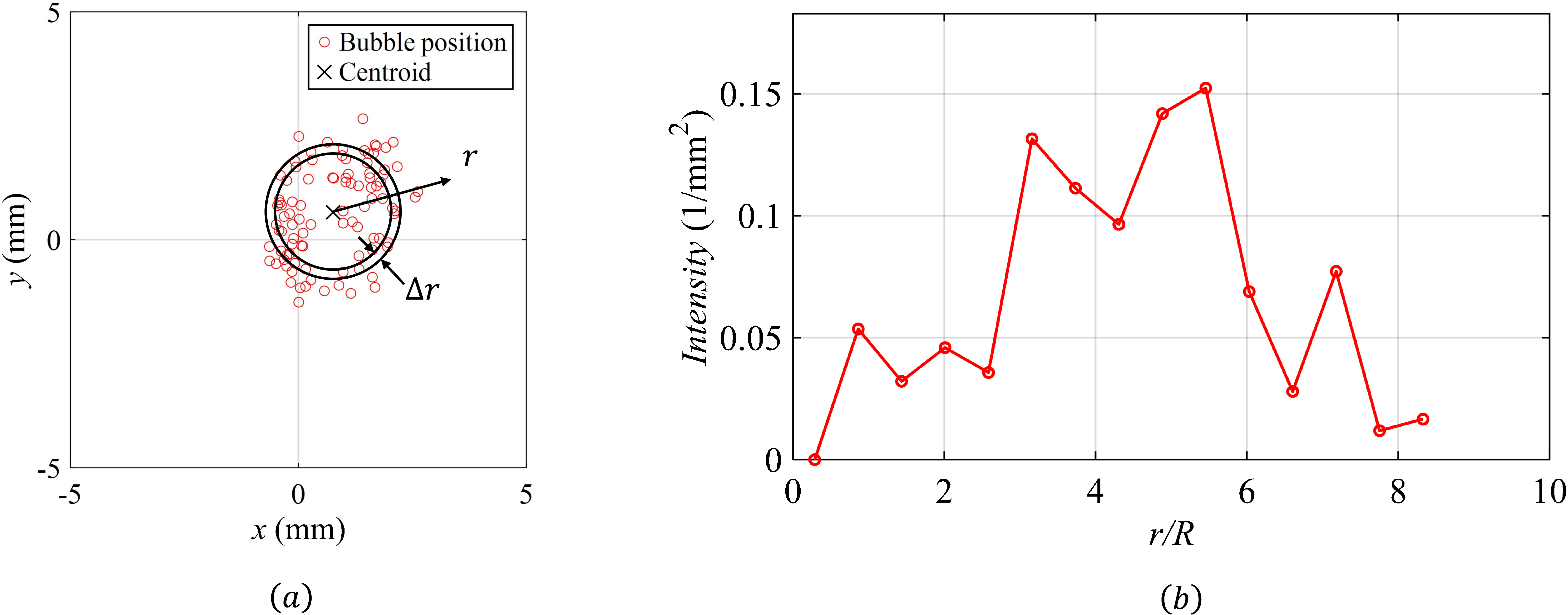}
\caption{\label{distribution} \added{(a) Bubble passage positions in the $x-y$ plane at $z= 140\,\mathrm{mm}$ and (b) radial number density of bubble passage positions with respect to the centroid.}}
\end{figure}

Figure~\ref{fig10} presents a schematic of the \replaced{characteristic dispersion structures at each cross section obtained from the experimental results}{three-dimensional dispersion structure observed experimentally} at $f= 12$, $20\,\mathrm{Hz}$ \added{in Fig.~\ref{fig5}}.
The red regions indicate areas where bubble passage positions concentrate at each \replaced{cross section}{height}. 
We divide the dispersion mechanism into two stages. In the first stage ( $40\,\mathrm{mm}\le z\le 80\,\mathrm{mm}$), the stable chained bubbles separate into two distinct streams. 
In the second stage ($100\,\mathrm{mm}\le z\le 140\,\mathrm{mm}$), individual bubbles exhibit significant positional variability and expand the elliptical dispersion region, with bubble passage positions accumulating near the edge of the dispersion region. 
\added{After the second stage, the bubble chain accumulation at the edge of the dispersion region can be characterized by analyzing the experimental spatial distribution of bubble passages. 
  Figure~\ref{distribution}(a) shows the bubble passage locations in a horizontal cross-section at $z= 140\,\mathrm{mm}$ from the nozzle, while Fig.~\ref{distribution}(b) presents the corresponding radial distributions normalized by the bubble radius $R$.
}
\added{These results correspond to experimental conditions of $f= 20\,\mathrm{Hz}$ and $d= 0.5\,\mathrm{mm}$. 
  The density distribution was calculated as follows:
}

\begin{equation}
\added{I(r) = \frac{N(r)}{N_{\text{total}}} \cdot \frac{1}{2\pi r \Delta r},}
\label{eq9}
\end{equation}
\added{where $N(r)$, i.e., the number of bubbles within the concentric annular region, is divided radially by $\Delta r\approx0.14\,\mathrm{mm}$, and divided by the total number of bubbles $N_{total}= 96$, normalizing by the area of each concentric region.}

\added{As shown, the distribution peaks within the range $3< r/R< 6$, indicating that, during the second stage of dispersion, bubbles are not uniformly distributed within the dispersion region; instead, they show reduced density near the center and a distinct tendency to accumulate along the outer edges. 
A similar trend is observed at $f= 12\,\mathrm{Hz}$.
}
We hypothesize that distinct physical mechanisms govern these two stages. 
We attribute the first stage to bubble--bubble interactions and relate the second stage to the upward flow generated by the bubbles rising through the liquid \added{(Right side of Fig.~\ref{fig10})}.

\vspace{\baselineskip}
\added{FIG. 11. in the original manuscript "Schematic of the two-way coupling between upward flow and bubbles rising in a chain." was removed.}
\vspace{\baselineskip}

\deleted{Figure~\ref{fig11} outlines the coupling between the liquid flow and bubbles rising in a chain.}
\iffalse
\begin{figure}%[t]
\includegraphics[width=0.5\linewidth]{Fig11}% Here is how to import EPS art
\caption{\label{fig11} Schematic of the two-way coupling between upward flow and bubbles rising in a chain. }
\end{figure} 
\fi
As a bubble rises through the liquid, it generates a wake and entrains the surrounding fluid \cite{Sanada2005}. 
Chained bubbles repeatedly pass through the liquid, inducing upward flow. 
The shear of upward flow generates a shear-induced lift force on each bubble, driving it away from the central axis. 
We therefore attribute the second-stage dispersion to shear-induced lift arising from the upward flow. 
We further expect that the superposition of individual bubble wakes generates this upward flow. 
\added{Even at a relatively large separation distance of $S\approx 20$, a bubble wake retains approximately $10\%$ of the bubble rise velocity (Fig. 16 in Hallez and Legendre \cite{Hallez2011}). 
  In the present experiments, the bubble chains at $f= 12$ and $20\,\mathrm{Hz}$ ($S = 25.3$ and $16.0$, respectively) likely generated a sustained upward flow, as subsequent bubbles passed through before the preceding wakes could completely attenuate. 
  In contrast, for $f= 4$ and $8\,\mathrm{Hz}$ ($S= 88.7$ and $30.3$, respectively), the spacing between successive bubbles was sufficient for the preceding wakes to fully dissipate. 
  As a result, persistent upward flow was not established, leading to a weaker secondary dispersion mechanism.
}
As bubbles disperse, the induced flow profile is modified, which feeds back into the promotion of further dispersion. 
This mutual interaction establishes a two-way coupling between bubble motion and liquid flow, thereby giving rise to the second-stage dispersion.

\subsection{Dispersion induced by bubble--bubble interactions}

\begin{figure}%[t]
\includegraphics[width=0.7\linewidth]{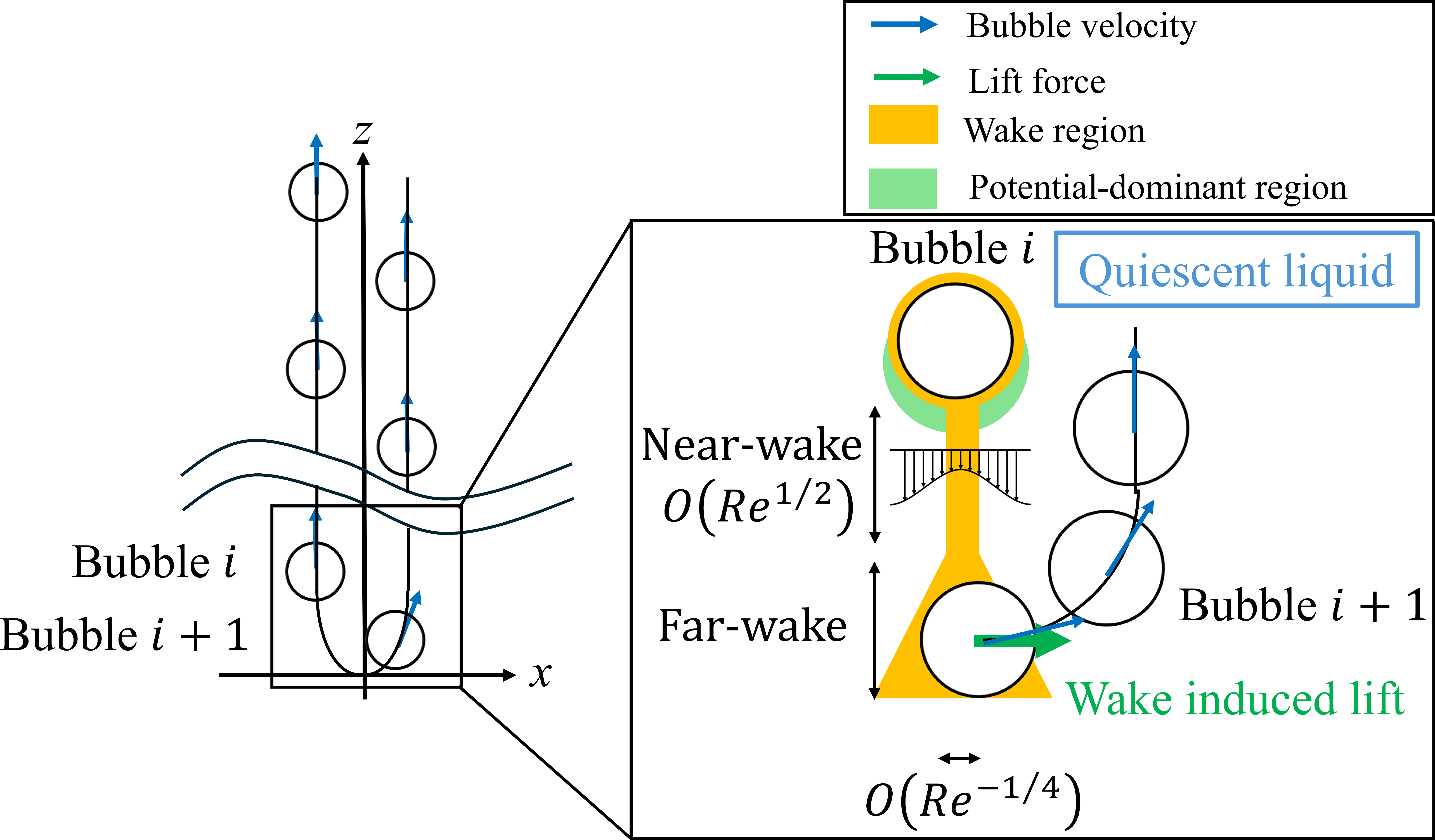}% Here is how to import EPS art
\caption{\label{fig12} Schematic of the formation mechanism of a U-shaped dispersion, with generated bubbles experiencing wake-induced lift that causes lateral migration, followed by bubbles rising vertically after exiting the wake region of a leading bubble.}
\end{figure}

The model including only bubble--bubble interactions reproduced the first-stage dispersion (separating into two steams), but did not reproduce the second stage dispersion (expanding the dispersion region) (Fig.~\ref{fig7}). 
Therefore, we conclude that bubble--bubble interaction governs the first-stage dispersion. 
Figure~\ref{fig12} presents the formation process for the small-scale U-shaped dispersion induced by bubble--bubble interaction. 
A trailing bubble experiences wake-induced lift within the wake region of a leading bubble and therefore migrates laterally, initiating dispersion. 
Lift force is generated by the far-wake of the leading bubble, even when the bubble separation $S$ exceeds the wake length of $O\left({Re}^{1/2}\right)$ \cite{RamirezMunoz2011a, RamirezMunoz2011b, RamirezMunoz2013}. 
The lateral migration of the bubbles rapidly decays after leaving the wake region, transitioning to vertical rising to form the U-shaped dispersion. 
\added{In the model, the bubbles rise linearly because only interactions between adjacent bubbles are considered. 
  During the initial dispersion stage, successive bubbles alternately separate; as the spacing between the bubbles increase, the interaction forces rapidly become negligible. 
  Consequently, each bubble subsequently rises with an approximately linear trajectory, behaving like an isolated single bubble.
}
Note that randomness of the initial bubble position was introduced to trigger dispersion in the model and does not influence the magnitude of dispersion (see the Appendix).
\added{Both the experiment and the model show that, during the first stage, the bubbles alternately separate into two streams. 
  When a given bubble experiences lift and deviates from the axis, the following bubble enters the wake of the preceding bubble and experiences a lift force in the opposite direction. 
  This recurring mechanism causes successive bubbles to deflect alternately in two directions. 
  Consequently, the first-stage dispersion is confined to a single vertical plane.
}

The dispersion width shows little frequency dependence for $f\ge 8\,\mathrm{Hz}$. 
The wake-induced dispersion is not simply governed by the gradual attenuation of bubble--bubble interaction with increasing bubble separation, but by whether the trailing bubble remains inside or outside the leading bubble wake region. 
The wake of a spherical bubble has a slender structure with less surface vorticity and no standing eddy. 
The bubble wake is maintained over long distances by forming a far-wake in more distant regions \cite{RamirezMunoz2011a, RamirezMunoz2011b, RamirezMunoz2013}. 
If the trailing bubble remains within the far-wake, the wake-induced lift acting on it remains nearly constant. 
In contrast, at $f=4\,\mathrm{Hz}$, the bubble separation exceeds the far-wake length. 
Therefore, there is no wake-induced and the bubbles rise vertically without dispersing.

\subsection{Dispersion induced by upward flow}

\begin{figure}%[t]
\includegraphics[width=0.7\linewidth]{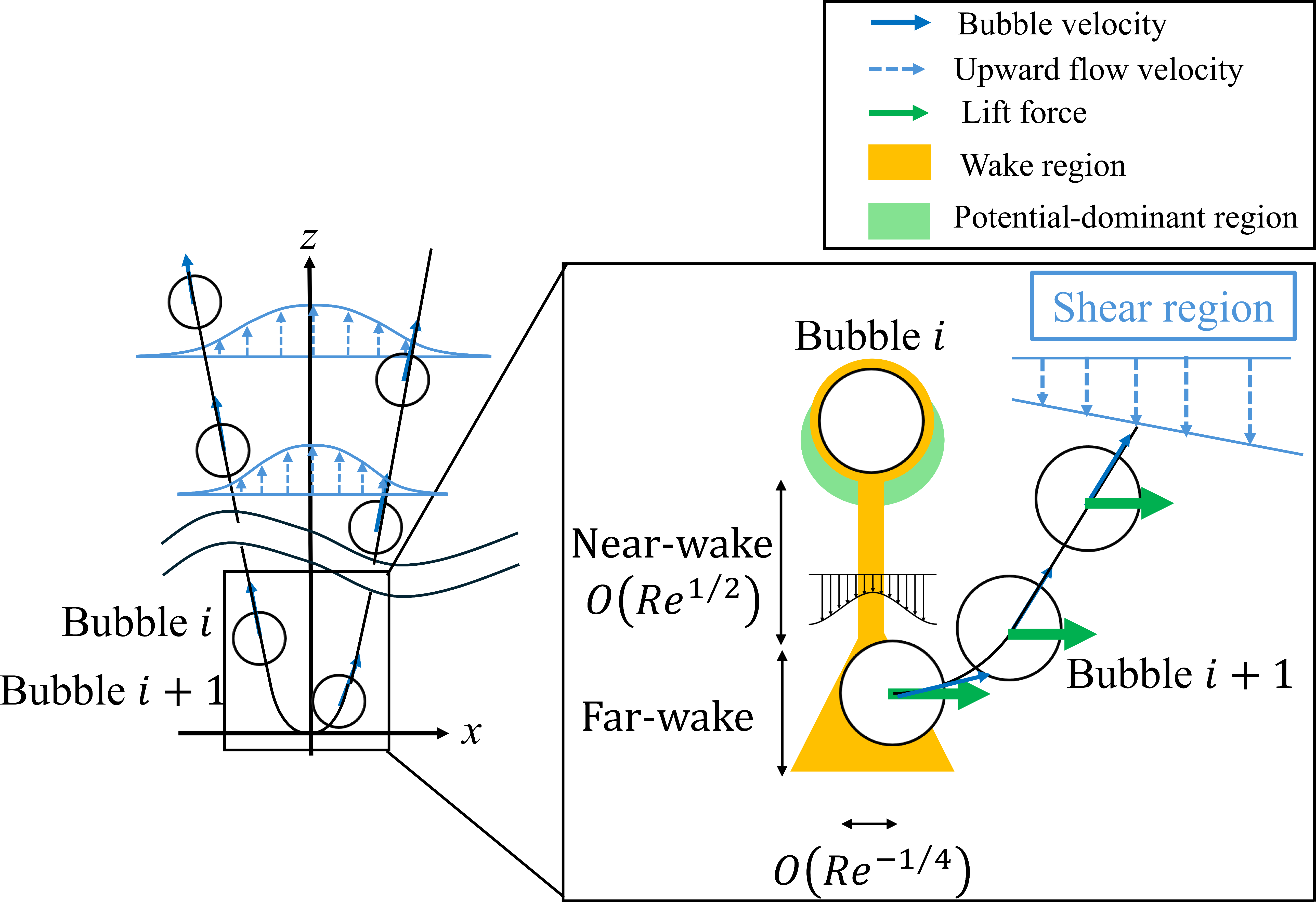}% Here is how to import EPS art
\caption{\label{fig13} Formation mechanism of a V-shaped dispersion. Generated bubbles experience wake-induced lift, causing lateral migration. After exiting the wake region of the leading bubble, bubbles experience shear-induced lift and rise obliquely.}
\end{figure}

Figure~\ref{fig13} shows the mechanism for dispersion driven by the upward flow. 
The lift force induced by the far-wake of the leading bubble causes the first stage dispersion. 
Once a bubble exits the wake region of the leading bubble, it experiences lift induced by the shear of the upward flow. 
The bubble continues to experience this lift as the upward flow expands, resulting in an increase in dispersion.
\added{Furthermore, the second stage dispersion driven by the upward flow is triggered by the initial displacement caused by the first stage dispersion. 
  The upward flow generates lift predominantly in the outer shear region, whereas this lift remains weak near the central axis. 
  Therefore, bubble--bubble interactions displace individual bubbles from the central axis to the outer shear region, where they experience lift from the upward flow.
}

The model presented herein constrains bubble motion to two dimensions. 
Experimentally, the upward flow likely exhibits a three-dimensional structure with a maximum velocity along the central axis. 
In three dimensions, the shear of the upward flow is expected to cause the bubbles to migrate laterally in a radial direction with respect to the central axis. 
The formation of a large-scale dispersion structure and the accumulation of bubbles near the edges arise from the upward flow.

At $f=4\,\mathrm{Hz}$, although chained bubbles show little dispersion from bubble--bubble interactions (Fig.~\ref{fig7}), dispersion was observed upon introducing the upward flow (Fig.~\ref{fig9}). 
This is attributed to the initial randomization of the $x-$position in the model, where the slight deviation from the central axis causes the bubble to experience lift from the upward flow. 
The detailed structure of the upward flow requires further investigation.

\subsection{Frequency dependence of the dispersion}

The bubble chain model demonstrates that the first-stage dispersion (separation into two streams) arises from bubble--bubble interaction, whereas the second-stage dispersion (expansion of the dispersion region) arises from the upward flow. 
Furthermore, the frequency dependence arises from the effect of the upward flow. 
We quantitatively evaluate the frequency dependence of the effect of bubble--bubble interaction and upward flow effect from experimentally observed bubbles. 

\begin{figure}%[t]
\includegraphics[width=0.4\linewidth]{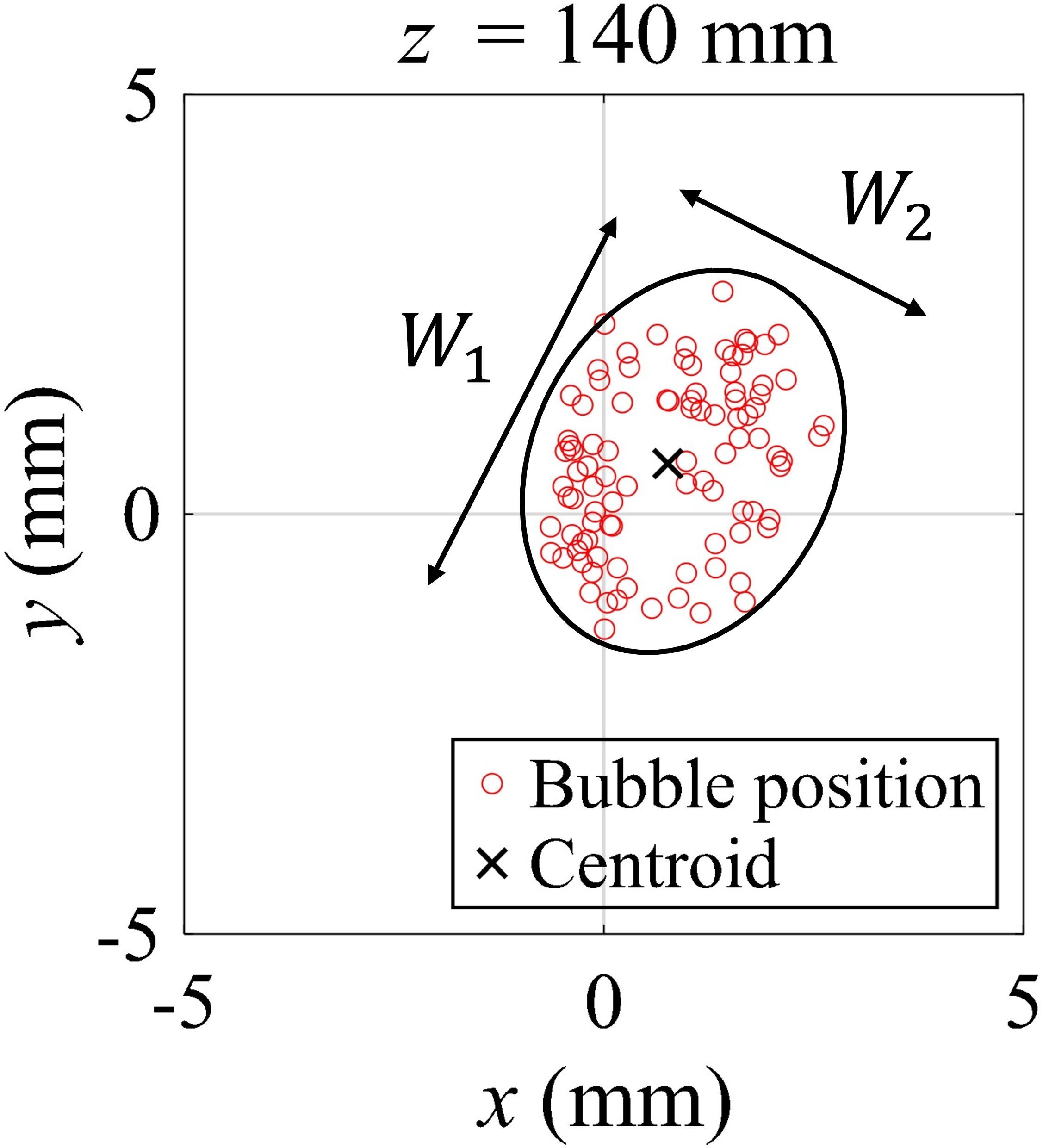}% Here is how to import EPS art
\caption{\label{fig14} Definition of the geometric dispersion parameters $W_1$ and $W_2$ for a cross-section of the bubble chain.}
\end{figure}

We introduce two geometric parameters to characterize the dispersion: the major-axis width $W_1$ and the minor-axis width $W_2$ (Fig.~\ref{fig14}).
We approximate the bubble passage distribution as an ellipse for each horizontal cross-section, and compute $W_1$ and $W_2$ using principal component analysis. 
We project the bubble passage coordinates ($x_i$, $y_i$) for each cross-section onto the first principal component and define $W_1$ as the difference between the maximum and minimum projected values; $W_2$ is analogously defined using the second principal component. 
\added{We determine the $95\%$ confidence intervals along the directions of the first and second principal components, defining their respective widths as $W_1$ and $W_2$. 
  For $f= 4\,\mathrm{Hz}$, we do not define the $95\%$ confidence intervals as there are a limited number of bubbles; instead, we directly employ the difference between the maximum and minimum values as $W_1$ and $W_2$.
}
According to the model analysis, the first stage of dispersion results from bubble--bubble interaction, and the second stage results from the upward flow. 
According to the experimental results, The bubbles spread in a particular direction during the first dispersion stage, increasing $W_1$ while leaving $W_2$ nearly unchanged. 
Bubbles disperse in radial directions during the second dispersion stage, increasing both $W_1$ and $W_2$. 
Therefore, $W_1$ reflects the combined effects of bubble--bubble interaction and upward flow, whereas $W_2$ isolates the upward flow contribution. 
Consequently, $W_1-W_2$ quantifies dispersion induced solely by bubble--bubble interaction.

\begin{figure}%[t]
\includegraphics[width=1\linewidth]{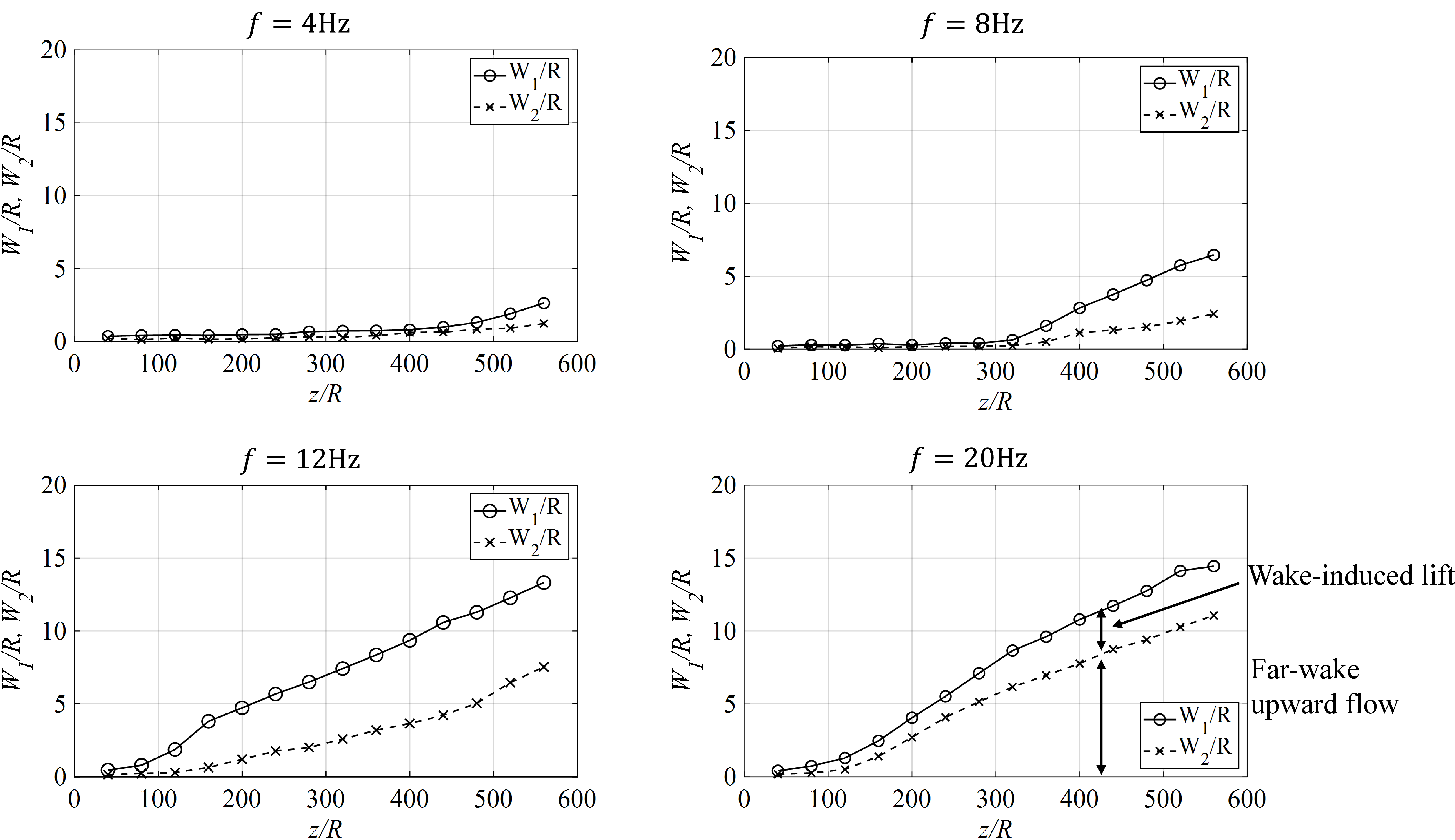}% Here is how to import EPS art
\caption{\label{fig15} \added{Relationship between the geometric dispersion parameters $W_1$ and $W_2$ and the height $z$.}}
\end{figure}

Figure~\ref{fig15} shows the relationship between dispersion and height $z$ for each frequency. 
The vertical axis shows the dimensionless dispersion parameters $W_1/R$ and $W_2/R$, and the horizontal axis shows the dimensionless height $z/R$, where $R$ is the bubble radius. 
For $f=4, 8\,\mathrm{Hz}$, both $W_1$ and $W_2$ remain nearly constant up to $\sim z/R=400$ and $300$, respectively. 
Above these heights, both parameters increase monotonically with $z/R$, indicating weak but sustained dispersion. Both $W_1-W_2$ (bubble--bubble interaction) and $W_2$ (upward flow) gradually increase, suggesting that modest bubble--bubble interaction combined with weak upward flow produces limited overall dispersion at these frequencies. 
For $f=12\,\mathrm{Hz}$, $W_1-W_2$ increases sharply at around $z/R=160$, before maintaining an approximately constant width thereafter. 
In contrast, the upward flow contribution $W_2$ increases monotonically with height. 
Thus, bubble--bubble interactions act locally at an early stage, whereas upward flow continuously enhances dispersion during bubble ascent. 
For $f=20\,\mathrm{Hz}$, $W_1-W_2$ gradually increases near $z/R=160$, then saturates and maintains a constant width. 
Meanwhile, $W_2$ continues to increase monotonically, indicating persistent dispersion driven by the upward flow.
These trends suggest that bubble--bubble interactions trigger dispersion locally at the early stage for higher frequencies ($f=12, 20\,\mathrm{Hz}$), while sustained lift generated by the upward flow dominates the subsequent dispersion region expansion.

\begin{figure}%[t]
\includegraphics[width=1\linewidth]{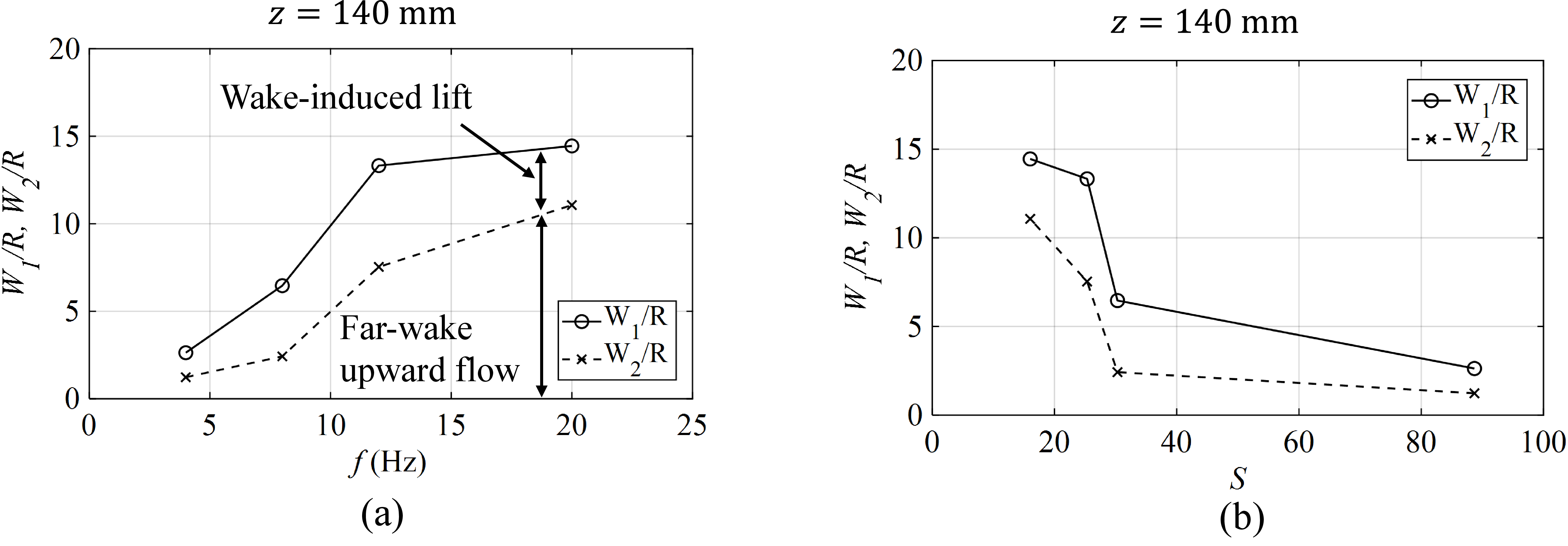}% Here is how to import EPS art
\caption{\label{fig16} \added{Relationship between dispersion parameter $W_1$ and $W_2$ and (a) generation frequency $f$ and (b) bubble separation $S$.}}
\end{figure}

Figure~\ref{fig16}(a) shows $W_1$ and $W_2$ at $z=140\,\mathrm{mm}$ as functions of $f$.
$W_1-W_2$  remains nearly independent of frequency, whereas $W_2$  increases with $f$. 
This trend agrees with the model prediction that the dispersion width for $f\ge 8\,\mathrm{Hz}$ does not have a strong frequency dependence when only bubble--bubble interactions occur. 
Therefore, the frequency dependence arises from structural changes in the induced upward flow rather than enhanced bubble--bubble interactions.

Because the bubble generation frequency $f$ controls the mean dimensionless bubble separation $S$, we replot the dispersion parameters $W_1$ and $W_2$ against $S$ in Fig.~\ref{fig16}(b). 
For $S< 30$, both $W_1$ and $W_2$ decrease sharply as $S$ increases. 
We infer that the upward flow primarily governs dispersion because $W_1-W_2$ shows little variation in this range. 
Both the upward flow contribution $W_2$ and the bubble--bubble contribution $W_1-W_2$ decrease for $S> 30$. 
The reduction in bubble--bubble interactions reflects the bubble separation exceeding the far-wake length. 
Although the wake of a spherical bubble exhibits a slender structure, its effective influence likely extends over a dimensionless length of approximately $30-89$.

\section{Conclusions}

We investigated the phenomenon of dispersion of bubbles rising in a chain through experiments and model-based predictions. 
We demonstrated that the large-scale dispersion structures in bubbles rising in a chain originate from bubble--bubble interactions and subsequently develop under the influence of induced upward flow. 
Experimentally, we observed three-dimensional dispersion of spherical bubbles produced by a originally-developed bubble generator at different frequencies. 
At $f=12, 20\,\mathrm{Hz}$, the dispersion occurred in two stages: concentration at two distinct streams by $z=80\,\mathrm{mm}$ and expansion of the dispersion region for $z>80\,\mathrm{mm}$. 
The bubble trajectories formed an elliptical dispersion structure, accumulating near the edge of the dispersion region. 
The formation of large-scale dispersion structures and frequency dependence are universal phenomena observed in clean spherical bubbles rising in a chain with different diameters ($d= 0.4, 0.5,$ and $0.6\,\mathrm{mm}$) at intermediate Reynolds number. 
Bubbles rising in a chain formed a relatively large-scale V-shaped dispersion even if the chain was composed of spherical bubbles with limited wake regions and the bubble separation exceeds the Moore's wake length \cite{Moore1963}.

We developed a bubble chain model that includes neighboring bubble interactions to predict the motion of bubbles in a chain. 
Although the model reproduced the initial dispersion of bubbles in a chain at $Re=O\left(50\right)$, it predicted only small-scale U-shaped dispersions and could not predict the formation of large-scale V-shaped dispersion. 
This indicated that bubble--bubble interactions trigger the onset of bubble dispersion in a chain but do not sustain the continuous lift required for further development of a V-shaped dispersion.

Further dispersion of bubbles in a chain arises from the upward flow formed by the bubbles themselves. 
Chained bubbles repeatedly pass along nearly identical paths, generating an upward flow around the chain. 
The upward flow shear induces an additional lift force on the bubbles. 
We verified the effect of upward flow on dispersion by incorporating it into the model. 
Even a weak upward flow promoted the formation of large-scale dispersion structures with the expected frequency dependence. 
Therefore, bubble dispersion initiates through lateral bubble migration due to wake-induced lift and subsequently develops through shear-induced lift due to the upward flow formed by the rising bubbles. 
These results explain the dispersion mechanism of spherical bubbles in a chain but also provide an interesting case study demonstrating the limitations of extending pairwise bubble interactions to multibubble systems and highlighting the importance of the mean flow induced by the bubbles themselves in bubble flows.

The model simplifies the three-dimensional dispersion by constraining it into two-dimensional dispersion. 
Furthermore, although the model demonstrates that the upward flow influences bubble motion through shear-induced lift, it does not yet clarify how the rising bubbles themselves generate the upward flow field. 
Therefore, future work should examine the effects of bubble--bubble interactions on dispersion and the structure of the upward flow in fully three-dimensional configurations, and address the two-way coupling between bubble motion and liquid flow.

\begin{acknowledgments}
This work was supported by JSPS KAKENHI Grant Number JP24K00801.
\added{The authors would like to thank both reviewers for their valuable suggestions regarding the quantitative evaluation of the experimental results, which greatly improved the manuscript through the addition and revision of the figures.
The authors are grateful to Mr. Ren Suzuki for kindly providing the data of PIV measurement.
}
\end{acknowledgments}

\appendix*

\section{Initial condition dependence of the dispersion}

\added{We introduced initial randomness in the experiment and the model to trigger the first-stage bubble interactions. 
  As demonstrated by Hallez and Legendre \cite{Hallez2011}, perfectly aligned bubbles do not experience lift force; therefore, generating wake-induced lift requires an initial perturbation. 
  This section describes the initial randomness and its subsequent effects in the numerical model and the experimental setup.
}

\begin{figure}%[t]
\includegraphics[width=0.7\linewidth]{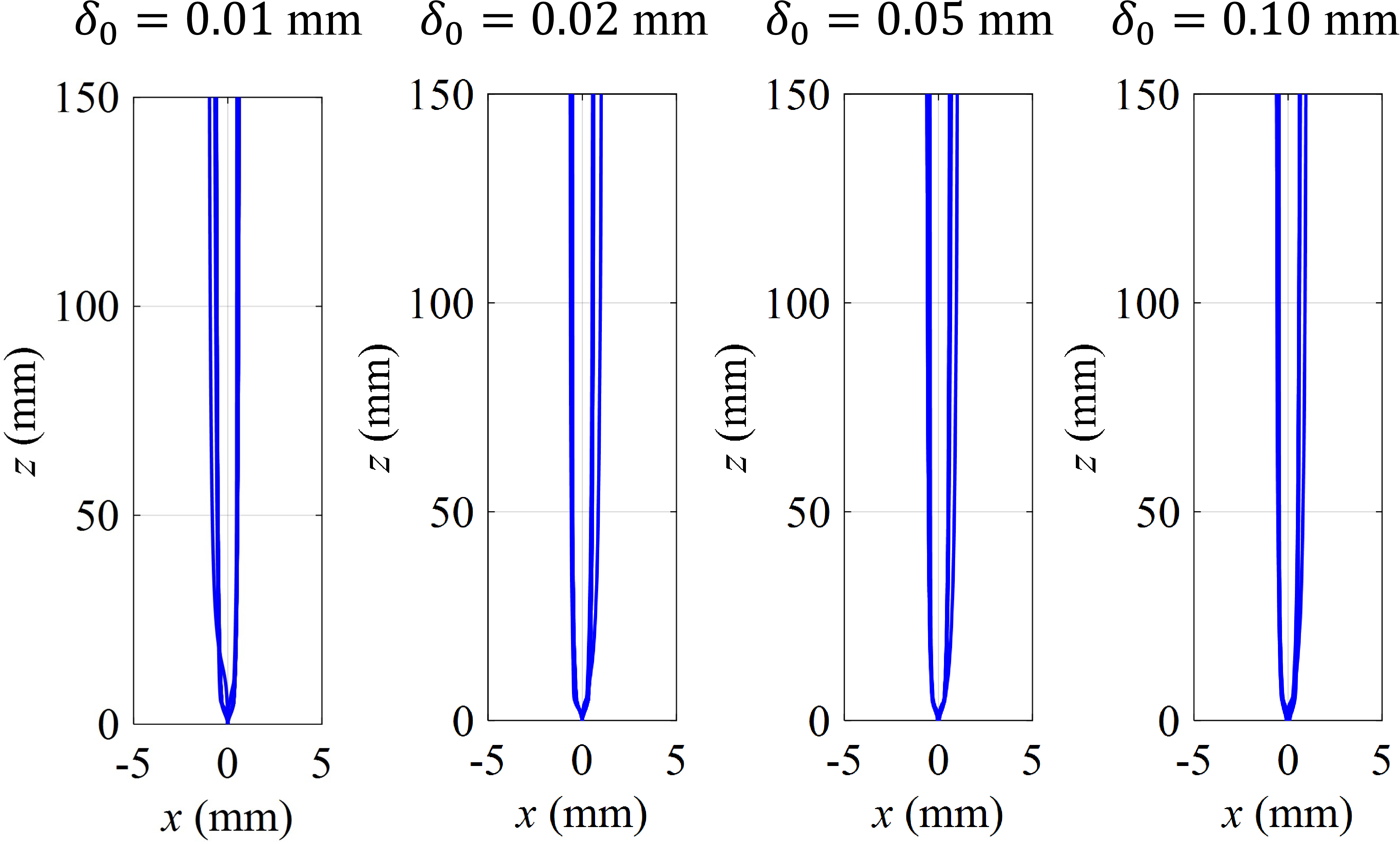}% Here is how to import EPS art
\caption{\label{fig17} The trajectories of bubbles in the model with increasing range of initial $x-$position randomness from left to right. ($f=20\,\mathrm{Hz}, d=0.5\,\mathrm{mm}, \nu=1 \,\mathrm{mm^2/s}$)}
\end{figure}

\added{In the model,} \replaced{randomness}{Randomness} was introduced to the model through the initial $x-$position of each bubble within the range $-0.005\le x_0\le 0.005\,\mathrm{mm}$ to induce dispersion through bubble--bubble interactions. 
\added{Therefore,} \replaced{it}{It} is therefore necessary to verify that the initial randomness acts only as the dispersion trigger and does not affect the subsequent bubble dynamics. 
Figure~\ref{fig17} 
shows the trajectories predicted by the bubble chain model when the initial range of the $x-$position $\delta_0$ varied ($\delta_0=0.01, 0.02, 0.05,$ and $0.10\,\mathrm{mm}$). 
\replaced{A change in the initial width does not display any}{No} significant change in the dispersion behavior \deleted{was observed as the initial width changed}, indicating that the randomness serves only as the dispersion trigger.

\begin{figure}%[t]
\includegraphics[width=1\linewidth]{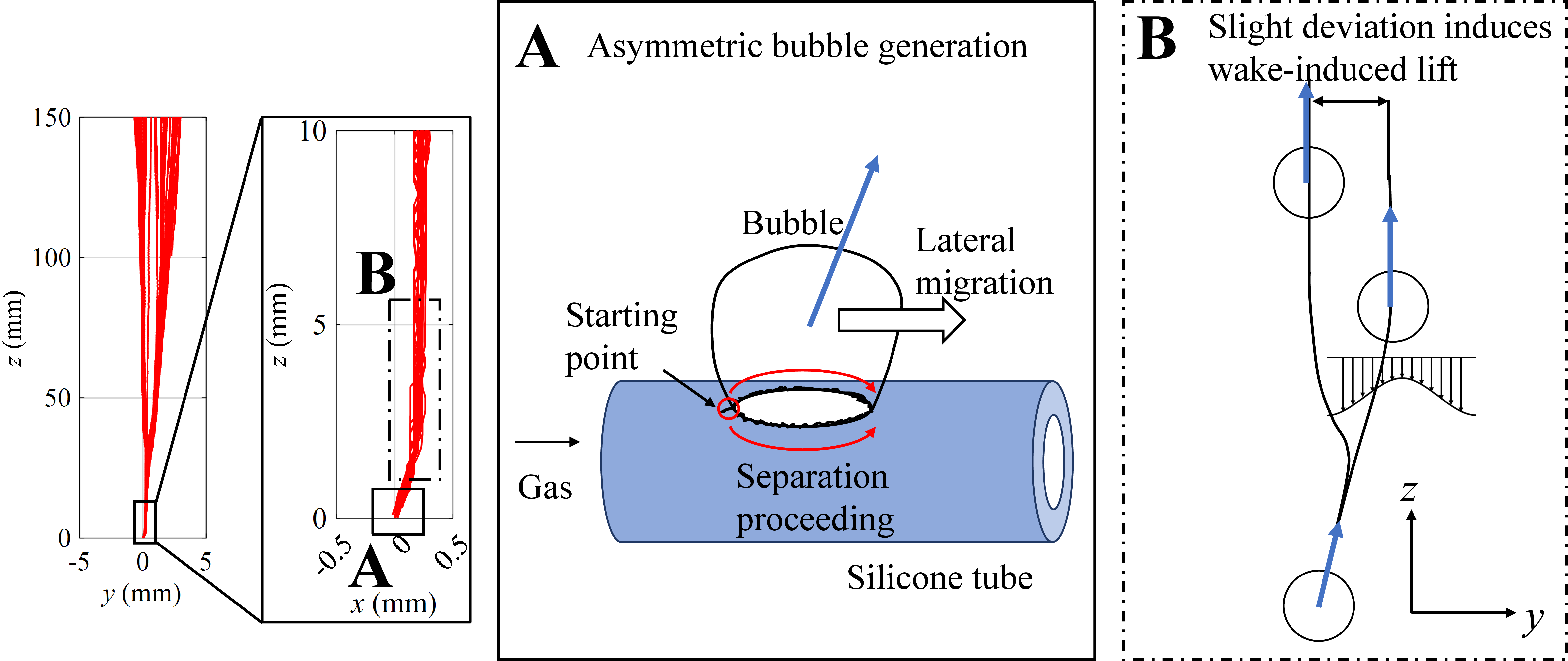}% Here is how to import EPS art
\caption{\label{fig18} (left) Trajectories of bubbles in a chain near the origin and (middle and right) schematic of the two-dimensional constrained dispersion process.}
\end{figure}

\replaced{
  In the experiments, the inherently three-dimensional nature of the flow implies that randomness not only triggers these interactions but also partially influences the orientation of the plane in which the primary dispersion develops. 
  This initial randomness appears as an asymmetric lateral displacement during bubble generation, originating from imperfections in the nozzle geometry.
}
{
The model reproduced the dispersion by constraining the system to two dimensions. 
However, in the real three-dimensional dispersion, the first stage of dispersion was constrained to a quasi-two-dimensional plane, as the trajectories separated into two streams. 
}
Figure~\ref{fig18} 
shows a magnified view of the bubble trajectories near the origin, corresponding to the experimental conditions $f=12\,\mathrm{Hz}$ and $d=0.5\,\mathrm{mm}$. 
Each bubble \added{slightly} shifts \deleted{slightly} to the right \replaced{following}{immediately after} generation and subsequently transitions to vertical ascent. 
\replaced{
  This behavior can be explained by the bubble detachment mechanism. 
  The nozzle used in this experiment has burrs along its edge to promote bubble detachment (Shirota \textit{et al.} \cite{Shirota2008}). 
  During generation, detachment initiates at a surface defect, causing the bubble to detach with a slight lateral displacement. 
  This initial offset subsequently triggers wake-induced lift. 
  Moreover, the direction of the first bubble determines the primary dispersion plane. 
  While the first bubble rises linearly like an isolated bubble, the second bubble experiences lift and deviates in a random lateral direction.
  The third bubble then experiences a maximum lift force in the opposite direction relative to the second bubble. 
  As a result, the trajectory of the second bubble sets the orientation of the dispersion plane, with the initial randomness influencing this direction.
}
{
This bubble behavior originates from asymmetric detachment from the orifice. 
The bubble detachment starts at a defect on the edge of the orifice, resulting in the bubble undergoing a small lateral displacement at the onset of upward motion. 
Because the detachment point reflects a geometric imperfection along the orifice rim, the lateral displacement depends on the specific orifice. 
Moreover, the initial variability in the lateral ejection is responsible for triggering the dispersion, with the directional bias promoting the experimentally observed quasi-two-dimensional dispersion. 
Based on these considerations, we conclude that the first stage of the bubble dispersion arises from bubble--bubble interactions.
}

\section{Particle image velocimetry of the upward flow induced by a bubble chain}

\begin{figure}%[t]
\includegraphics[width=0.6\linewidth]{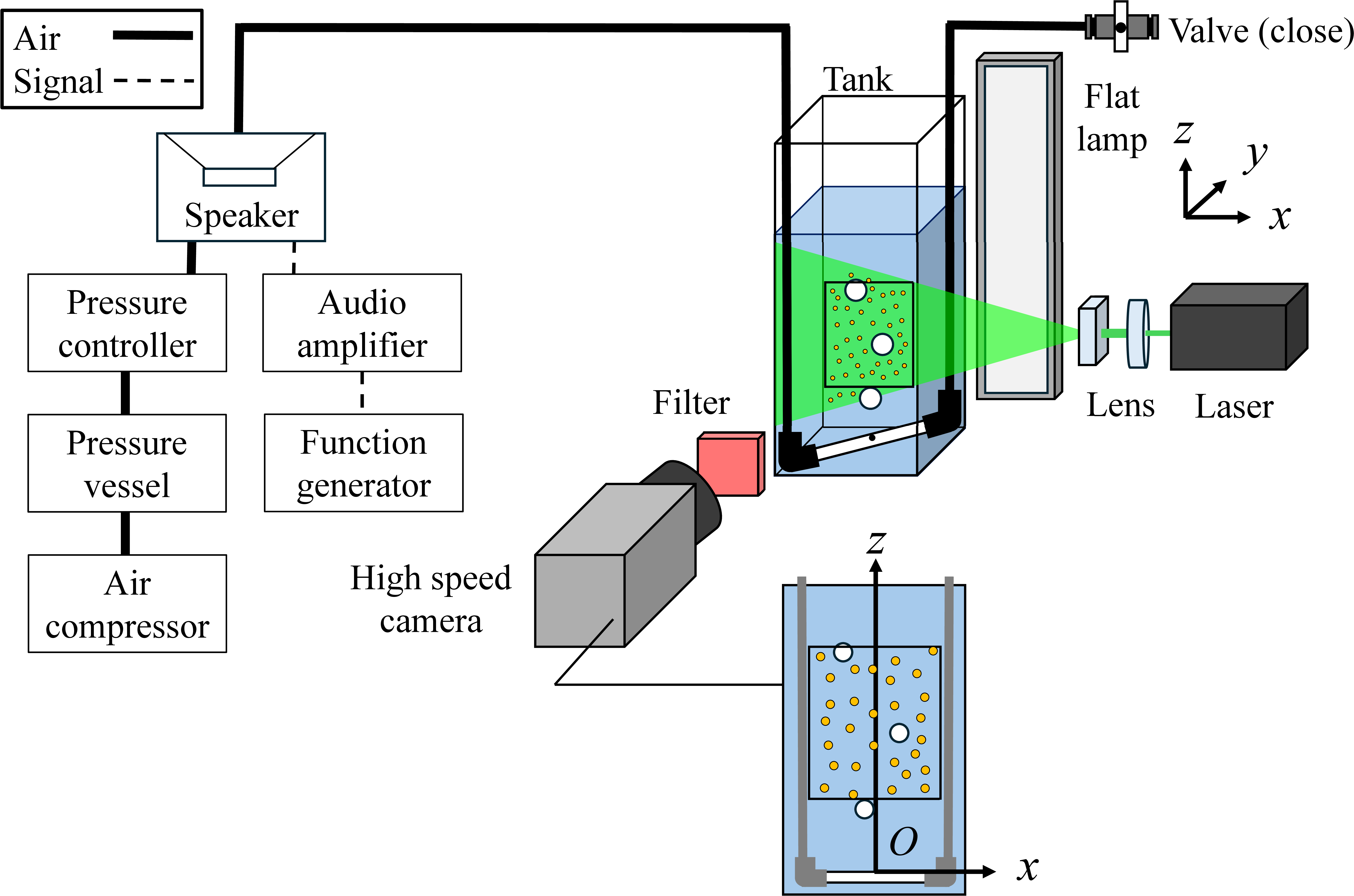}% Here is how to import EPS art
\caption{\label{PIVsetup}  \added{Schematic of the bubble generation system and particle image velocimetry measurement setup.}}
\end{figure}

\begin{figure}%[t]
\includegraphics[width=0.6\linewidth]{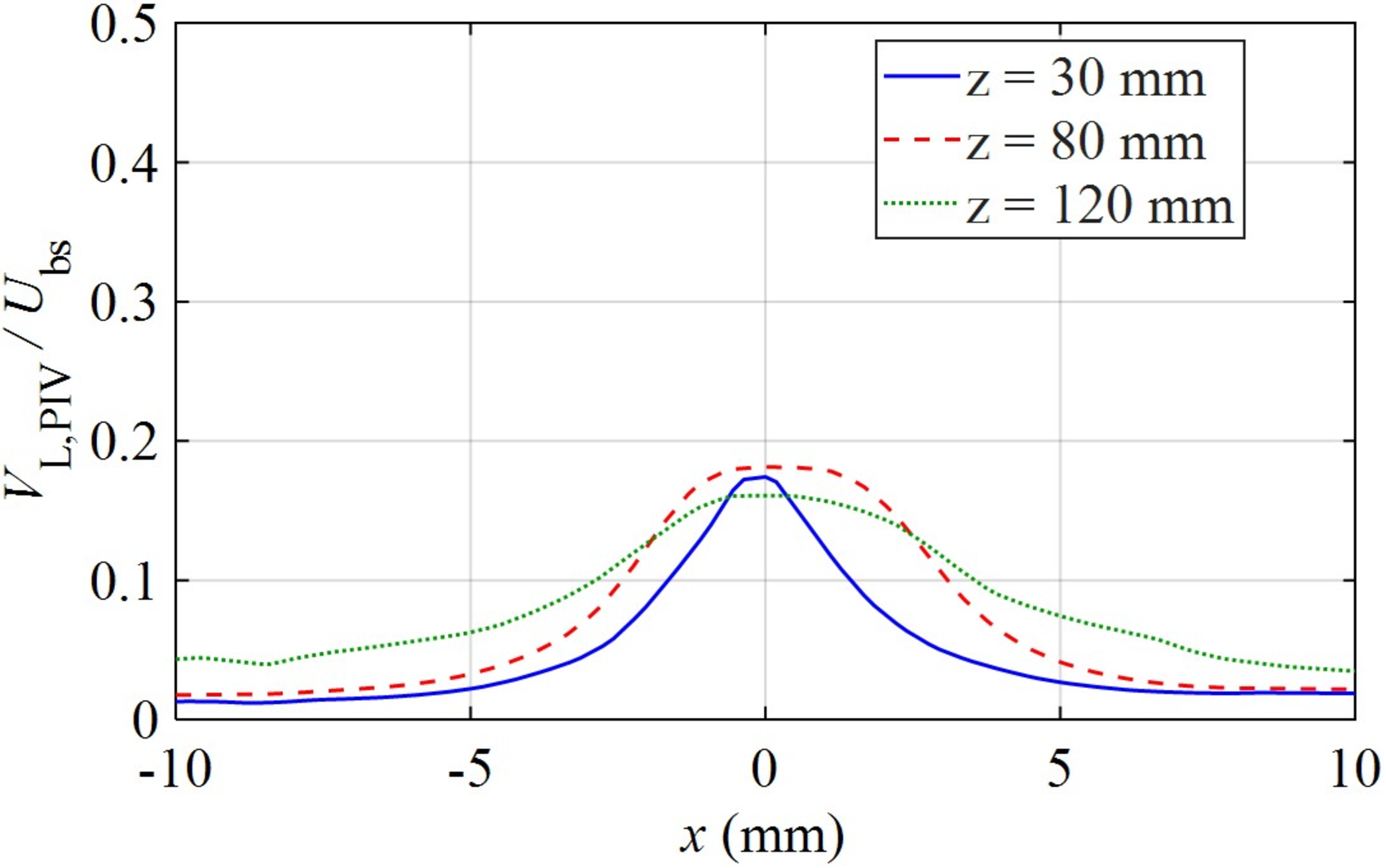}% Here is how to import EPS art
\caption{\label{PIVresult} \added{The upward flow velocity distribution measured by particle image velocimetry ($f= 20\,\mathrm{Hz}$, $d= 0.5\,\mathrm{mm}$, silicone oil ($\nu= 1\,\mathrm{mm^2/s}$)).}}
\end{figure}

\added{We experimentally verified the presence of the predicted upward flow using particle image velocimetry (PIV). 
  Figure~\ref{PIVsetup} shows the experimental setup (integrating the bubble generation system from Fig.~\ref{fig1}).
}
\added{To measure the velocity distribution in the $x-z$ plane, fluorescent polyethylene tracer particles containing rhodamine B were added to the silicone oil. 
  The measurement plane was illuminated by a laser sheet from the side, and particle motion was recorded via a long-pass filter. 
  Figure~\ref{PIVresult} presents representative measurements of the liquid velocity distribution obtained by PIV for a rising chain of clean, spherical bubbles with diameter $d= 0.5\,\mathrm{mm}$ at a frequency of $f= 20\,\mathrm{Hz}$.
}
\added{The bubbles are generated at $(x, z) = (0, 0)$, and the figure shows vertical velocity profiles at heights of $z= 30$, $80$, and $120\,\mathrm{mm}$. 
  The vertical velocity is non-dimensionalized by the terminal rise velocity of a single bubble, $U_{bs}$. 
  The profiles reveal a high-velocity region near the central axis, providing experimental confirmation of the upward flow. 
  The measured upward velocity remains within $20\%$ of $U_{bs}$, supporting the model validity introduced in Fig.~\ref{fig3}. 
  In addition, as the bubble chain disperses with increasing $z$, the upward flow spreads laterally, consistent with the trends shown in Fig.~\ref{fig3}. 
  The tracer particles are assumed to have a negligible effect on bubble dynamics in silicone oil (Zenit \textit{et al.} \cite{zenit2009measurements}).
}

\section{Bifurcation diagram of a clean spherical bubble chain}

\begin{figure}%[t]
\includegraphics[width=0.6\linewidth]{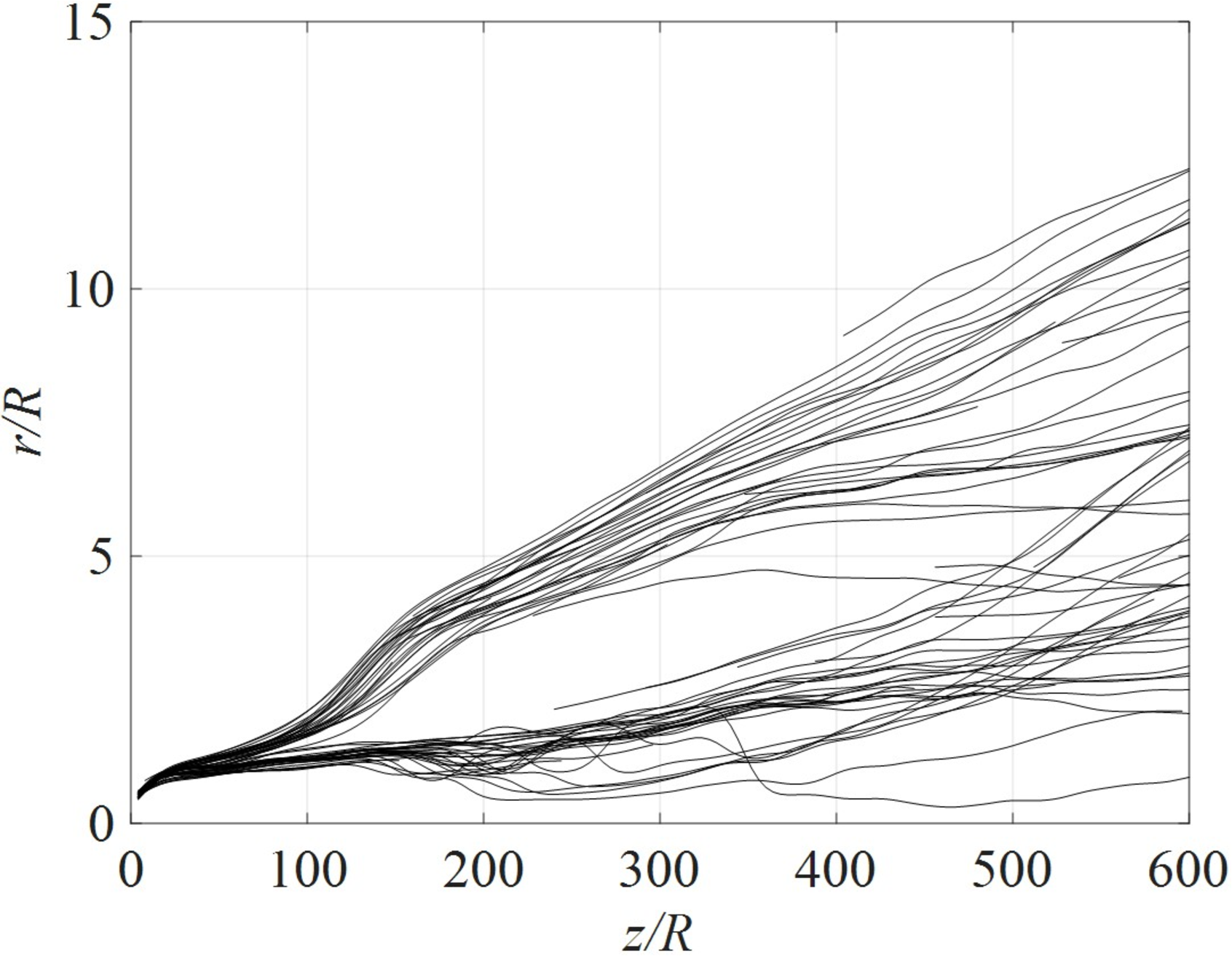}% Here is how to import EPS art
\caption{\label{bifurcation} \added{The bifurcation diagram of a clean spherical bubble chain ($f= 12\,\mathrm{Hz}$, $d= 0.5\,\mathrm{mm}$, silicone oil ($\nu= 1\,\mathrm{mm^2/s}$)).}}
\end{figure}

\added{To illustrate the nonlinear nature of bubble chain dispersion, we constructed a bifurcation diagram from experimental data at $f= 12\,\mathrm{Hz}$  and $d= 0.5\,\mathrm{mm}$ , as shown in Fig.~\ref{bifurcation}.
}
\added{The vertical axis represents the radial distance from the origin (the initial $x-$ and $y-$coordinates of bubble generation) taken as the characteristic variable, while the horizontal axis denotes the downstream height from the nozzle, used here as the control parameter. 
  Both quantities are nondimensionalized by the bubble radius $R$. 
  After generation, each bubble rises with a slight lateral displacement from the origin, arising from the asymmetry in the detachment process described earlier. 
  A clear bifurcation occurs at $z/R\approx 100$, corresponding to the onset of the first-stage dispersion. 
  Beyond this point, the trajectories exhibit increased spread, indicating the transition to the second-stage dispersion. 
  Because the radial distance from the origin is used as the characteristic variable, the apparent scattering of trajectories caused by the initial detachment asymmetry is clearly captured in Fig.~\ref{bifurcation}. 
  However, the real system is not purely chaotic; as demonstrated in Fig.~\ref{distribution}, bubbles in the second-stage dispersion form a distinctly nonuniform distribution, with a preferential accumulation near the outer edge of the dispersion region.
}

%\nocite{*}

\bibliography{apssamp}% Produces the bibliography via BibTeX.

\providecommand{\noopsort}[1]{}\providecommand{\singleletter}[1]{#1}%
\begin{thebibliography}{34}%
\makeatletter
\providecommand \@ifxundefined [1]{%
 \@ifx{#1\undefined}
}%
\providecommand \@ifnum [1]{%
 \ifnum #1\expandafter \@firstoftwo
 \else \expandafter \@secondoftwo
 \fi
}%
\providecommand \@ifx [1]{%
 \ifx #1\expandafter \@firstoftwo
 \else \expandafter \@secondoftwo
 \fi
}%
\providecommand \natexlab [1]{#1}%
\providecommand \enquote  [1]{``#1''}%
\providecommand \bibnamefont  [1]{#1}%
\providecommand \bibfnamefont [1]{#1}%
\providecommand \citenamefont [1]{#1}%
\providecommand \href@noop [0]{\@secondoftwo}%
\providecommand \href [0]{\begingroup \@sanitize@url \@href}%
\providecommand \@href[1]{\@@startlink{#1}\@@href}%
\providecommand \@@href[1]{\endgroup#1\@@endlink}%
\providecommand \@sanitize@url [0]{\catcode `\\12\catcode `\$12\catcode `\&12\catcode `\#12\catcode `\^12\catcode `\_12\catcode `\%12\relax}%
\providecommand \@@startlink[1]{}%
\providecommand \@@endlink[0]{}%
\providecommand \url  [0]{\begingroup\@sanitize@url \@url }%
\providecommand \@url [1]{\endgroup\@href {#1}{\urlprefix }}%
\providecommand \urlprefix  [0]{URL }%
\providecommand \Eprint [0]{\href }%
\providecommand \doibase [0]{https://doi.org/}%
\providecommand \selectlanguage [0]{\@gobble}%
\providecommand \bibinfo  [0]{\@secondoftwo}%
\providecommand \bibfield  [0]{\@secondoftwo}%
\providecommand \translation [1]{[#1]}%
\providecommand \BibitemOpen [0]{}%
\providecommand \bibitemStop [0]{}%
\providecommand \bibitemNoStop [0]{.\EOS\space}%
\providecommand \EOS [0]{\spacefactor3000\relax}%
\providecommand \BibitemShut  [1]{\csname bibitem#1\endcsname}%
\let\auto@bib@innerbib\@empty
%</preamble>
\bibitem [{\citenamefont {Ruzicka}(2000)}]{Ruzicka2000}%
  \BibitemOpen
  \bibfield  {author} {\bibinfo {author} {\bibfnamefont {M.~C.}\ \bibnamefont {Ruzicka}},\ }\bibfield  {title} {\bibinfo {title} {On bubbles rising in line},\ }\href@noop {} {\bibfield  {journal} {\bibinfo  {journal} {Int. J. Multiphase Flow}\ }\textbf {\bibinfo {volume} {26}},\ \bibinfo {pages} {1141} (\bibinfo {year} {2000})}\BibitemShut {NoStop}%
\bibitem [{\citenamefont {Sanada}\ \emph {et~al.}(2005)\citenamefont {Sanada}, \citenamefont {Watanabe}, \citenamefont {Fukano},\ and\ \citenamefont {Kariyasaki}}]{Sanada2005}%
  \BibitemOpen
  \bibfield  {author} {\bibinfo {author} {\bibfnamefont {T.}~\bibnamefont {Sanada}}, \bibinfo {author} {\bibfnamefont {M.}~\bibnamefont {Watanabe}}, \bibinfo {author} {\bibfnamefont {T.}~\bibnamefont {Fukano}},\ and\ \bibinfo {author} {\bibfnamefont {A.}~\bibnamefont {Kariyasaki}},\ }\bibfield  {title} {\bibinfo {title} {Behavior of a single coherent gas bubble chain and surrounding liquid jet flow structure},\ }\href@noop {} {\bibfield  {journal} {\bibinfo  {journal} {Chem. Eng. Sci.}\ }\textbf {\bibinfo {volume} {60}},\ \bibinfo {pages} {4886} (\bibinfo {year} {2005})}\BibitemShut {NoStop}%
\bibitem [{\citenamefont {Atasi}\ \emph {et~al.}(2023)\citenamefont {Atasi}, \citenamefont {Ravisankar}, \citenamefont {Legendre},\ and\ \citenamefont {Zenit}}]{Atasi2023}%
  \BibitemOpen
  \bibfield  {author} {\bibinfo {author} {\bibfnamefont {O.}~\bibnamefont {Atasi}}, \bibinfo {author} {\bibfnamefont {M.}~\bibnamefont {Ravisankar}}, \bibinfo {author} {\bibfnamefont {D.}~\bibnamefont {Legendre}},\ and\ \bibinfo {author} {\bibfnamefont {R.}~\bibnamefont {Zenit}},\ }\bibfield  {title} {\bibinfo {title} {Presence of surfactants controls the stability of bubble chains in carbonated drinks},\ }\href@noop {} {\bibfield  {journal} {\bibinfo  {journal} {Phys. Rev. Fluids}\ }\textbf {\bibinfo {volume} {8}},\ \bibinfo {pages} {053601} (\bibinfo {year} {2023})}\BibitemShut {NoStop}%
\bibitem [{\citenamefont {Legendre}\ and\ \citenamefont {Zenit}(2025)}]{legendre2025gas}%
  \BibitemOpen
  \bibfield  {author} {\bibinfo {author} {\bibfnamefont {D.}~\bibnamefont {Legendre}}\ and\ \bibinfo {author} {\bibfnamefont {R.}~\bibnamefont {Zenit}},\ }\bibfield  {title} {\bibinfo {title} {Gas bubble dynamics},\ }\href@noop {} {\bibfield  {journal} {\bibinfo  {journal} {Rev. Mod. Phys.}\ }\textbf {\bibinfo {volume} {97}},\ \bibinfo {pages} {025001} (\bibinfo {year} {2025})}\BibitemShut {NoStop}%
\bibitem [{\citenamefont {Moore}(1963)}]{Moore1963}%
  \BibitemOpen
  \bibfield  {author} {\bibinfo {author} {\bibfnamefont {D.~W.}\ \bibnamefont {Moore}},\ }\bibfield  {title} {\bibinfo {title} {The boundary layer on a spherical gas bubble},\ }\href@noop {} {\bibfield  {journal} {\bibinfo  {journal} {J. Fluid Mech.}\ }\textbf {\bibinfo {volume} {16}},\ \bibinfo {pages} {161} (\bibinfo {year} {1963})}\BibitemShut {NoStop}%
\bibitem [{\citenamefont {Blanco}\ and\ \citenamefont {Magnaudet}(1995)}]{Blanco1995}%
  \BibitemOpen
  \bibfield  {author} {\bibinfo {author} {\bibfnamefont {A.}~\bibnamefont {Blanco}}\ and\ \bibinfo {author} {\bibfnamefont {J.}~\bibnamefont {Magnaudet}},\ }\bibfield  {title} {\bibinfo {title} {The structure of the axisymmetric high-reynolds number flow around an ellipsoidal bubble of fixed shape},\ }\href@noop {} {\bibfield  {journal} {\bibinfo  {journal} {Phys. Fluids}\ }\textbf {\bibinfo {volume} {7}},\ \bibinfo {pages} {1265} (\bibinfo {year} {1995})}\BibitemShut {NoStop}%
\bibitem [{\citenamefont {Harper}(1970)}]{Harper1970}%
  \BibitemOpen
  \bibfield  {author} {\bibinfo {author} {\bibfnamefont {J.~F.}\ \bibnamefont {Harper}},\ }\bibfield  {title} {\bibinfo {title} {On bubbles rising in line at large reynolds numbers},\ }\href@noop {} {\bibfield  {journal} {\bibinfo  {journal} {J. Fluid Mech.}\ }\textbf {\bibinfo {volume} {41}},\ \bibinfo {pages} {751} (\bibinfo {year} {1970})}\BibitemShut {NoStop}%
\bibitem [{\citenamefont {Yuan}\ and\ \citenamefont {Prosperetti}(1994)}]{Yuan1994}%
  \BibitemOpen
  \bibfield  {author} {\bibinfo {author} {\bibfnamefont {H.}~\bibnamefont {Yuan}}\ and\ \bibinfo {author} {\bibfnamefont {A.}~\bibnamefont {Prosperetti}},\ }\bibfield  {title} {\bibinfo {title} {On the in-line motion of two spherical bubbles in a viscous fluid},\ }\href@noop {} {\bibfield  {journal} {\bibinfo  {journal} {J. Fluid Mech.}\ }\textbf {\bibinfo {volume} {278}},\ \bibinfo {pages} {325} (\bibinfo {year} {1994})}\BibitemShut {NoStop}%
\bibitem [{\citenamefont {Harper}(1997)}]{Harper1997}%
  \BibitemOpen
  \bibfield  {author} {\bibinfo {author} {\bibfnamefont {J.~F.}\ \bibnamefont {Harper}},\ }\bibfield  {title} {\bibinfo {title} {Bubbles rising in line: why is the first approximation so bad?},\ }\href@noop {} {\bibfield  {journal} {\bibinfo  {journal} {J. Fluid Mech.}\ }\textbf {\bibinfo {volume} {351}},\ \bibinfo {pages} {289} (\bibinfo {year} {1997})}\BibitemShut {NoStop}%
\bibitem [{\citenamefont {Biesheuvel}\ and\ \citenamefont {Wijngaarden}(1982)}]{Biesheuvel1982}%
  \BibitemOpen
  \bibfield  {author} {\bibinfo {author} {\bibfnamefont {A.}~\bibnamefont {Biesheuvel}}\ and\ \bibinfo {author} {\bibfnamefont {L.~V.}\ \bibnamefont {Wijngaarden}},\ }\bibfield  {title} {\bibinfo {title} {The motion of pairs of gas bubbles in a perfect liquid},\ }\href@noop {} {\bibfield  {journal} {\bibinfo  {journal} {J. Eng. Math.}\ }\textbf {\bibinfo {volume} {16}},\ \bibinfo {pages} {349} (\bibinfo {year} {1982})}\BibitemShut {NoStop}%
\bibitem [{\citenamefont {Kok}(1993)}]{Kok1993}%
  \BibitemOpen
  \bibfield  {author} {\bibinfo {author} {\bibfnamefont {J.~B.~W.}\ \bibnamefont {Kok}},\ }\bibfield  {title} {\bibinfo {title} {Dynamics of a pair of gas bubbles moving through liquid. i: theory},\ }\href@noop {} {\bibfield  {journal} {\bibinfo  {journal} {Eur. J. Mech. B Fluids}\ }\textbf {\bibinfo {volume} {12}},\ \bibinfo {pages} {515} (\bibinfo {year} {1993})}\BibitemShut {NoStop}%
\bibitem [{\citenamefont {Levich}(1962)}]{Levich1962}%
  \BibitemOpen
  \bibfield  {author} {\bibinfo {author} {\bibfnamefont {V.~G.}\ \bibnamefont {Levich}},\ }\href@noop {} {\emph {\bibinfo {title} {Physicochemical Hydrodynamics}}}\ (\bibinfo  {publisher} {Prentice-Hall},\ \bibinfo {year} {1962})\BibitemShut {NoStop}%
\bibitem [{\citenamefont {Legendre}\ and\ \citenamefont {Magnaudet}(1998)}]{Legendre1998}%
  \BibitemOpen
  \bibfield  {author} {\bibinfo {author} {\bibfnamefont {D.}~\bibnamefont {Legendre}}\ and\ \bibinfo {author} {\bibfnamefont {J.}~\bibnamefont {Magnaudet}},\ }\bibfield  {title} {\bibinfo {title} {The lift force on a spherical bubble in a viscous linear shear flow},\ }\href@noop {} {\bibfield  {journal} {\bibinfo  {journal} {J. Fluid Mech.}\ }\textbf {\bibinfo {volume} {368}},\ \bibinfo {pages} {81} (\bibinfo {year} {1998})}\BibitemShut {NoStop}%
\bibitem [{\citenamefont {Tomiyama}\ \emph {et~al.}(2002)\citenamefont {Tomiyama}, \citenamefont {Tamai}, \citenamefont {Zun},\ and\ \citenamefont {Hosokawa}}]{Tomiyama2002}%
  \BibitemOpen
  \bibfield  {author} {\bibinfo {author} {\bibfnamefont {A.}~\bibnamefont {Tomiyama}}, \bibinfo {author} {\bibfnamefont {H.}~\bibnamefont {Tamai}}, \bibinfo {author} {\bibfnamefont {I.}~\bibnamefont {Zun}},\ and\ \bibinfo {author} {\bibfnamefont {S.}~\bibnamefont {Hosokawa}},\ }\bibfield  {title} {\bibinfo {title} {Transverse migration of single bubbles in simple shear flows},\ }\href@noop {} {\bibfield  {journal} {\bibinfo  {journal} {Chem. Eng. Sci.}\ }\textbf {\bibinfo {volume} {57}},\ \bibinfo {pages} {1849} (\bibinfo {year} {2002})}\BibitemShut {NoStop}%
\bibitem [{\citenamefont {Hallez}\ and\ \citenamefont {Legendre}(2011)}]{Hallez2011}%
  \BibitemOpen
  \bibfield  {author} {\bibinfo {author} {\bibfnamefont {Y.}~\bibnamefont {Hallez}}\ and\ \bibinfo {author} {\bibfnamefont {D.}~\bibnamefont {Legendre}},\ }\bibfield  {title} {\bibinfo {title} {Interaction between two spherical bubbles rising in a viscous liquid},\ }\href@noop {} {\bibfield  {journal} {\bibinfo  {journal} {J. Fluid Mech.}\ }\textbf {\bibinfo {volume} {673}},\ \bibinfo {pages} {406} (\bibinfo {year} {2011})}\BibitemShut {NoStop}%
\bibitem [{\citenamefont {Sangani}\ \emph {et~al.}(2001)\citenamefont {Sangani}, \citenamefont {Zenit},\ and\ \citenamefont {Koch}}]{Sangani2001}%
  \BibitemOpen
  \bibfield  {author} {\bibinfo {author} {\bibfnamefont {A.~S.}\ \bibnamefont {Sangani}}, \bibinfo {author} {\bibfnamefont {R.}~\bibnamefont {Zenit}},\ and\ \bibinfo {author} {\bibfnamefont {D.~L.}\ \bibnamefont {Koch}},\ }\bibfield  {title} {\bibinfo {title} {Measurements of the average properties of a suspension of bubbles rising in a vertical channel},\ }\href@noop {} {\bibfield  {journal} {\bibinfo  {journal} {J. Fluid Mech.}\ }\textbf {\bibinfo {volume} {429}},\ \bibinfo {pages} {307} (\bibinfo {year} {2001})}\BibitemShut {NoStop}%
\bibitem [{\citenamefont {Takagi}\ and\ \citenamefont {Matsumoto}(2011)}]{Takagi2011}%
  \BibitemOpen
  \bibfield  {author} {\bibinfo {author} {\bibfnamefont {S.}~\bibnamefont {Takagi}}\ and\ \bibinfo {author} {\bibfnamefont {Y.}~\bibnamefont {Matsumoto}},\ }\bibfield  {title} {\bibinfo {title} {Surfactant effects on bubble motion and bubbly flow},\ }\href@noop {} {\bibfield  {journal} {\bibinfo  {journal} {Annu. Rev. Fluid Mech.}\ }\textbf {\bibinfo {volume} {43}},\ \bibinfo {pages} {615} (\bibinfo {year} {2011})}\BibitemShut {NoStop}%
\bibitem [{\citenamefont {Lee}\ and\ \citenamefont {Choi}(2026)}]{Lee2026}%
  \BibitemOpen
  \bibfield  {author} {\bibinfo {author} {\bibfnamefont {I.}~\bibnamefont {Lee}}\ and\ \bibinfo {author} {\bibfnamefont {H.}~\bibnamefont {Choi}},\ }\bibfield  {title} {\bibinfo {title} {A numerical study on the clustering characteristics of rising air bubbles in stagnant water},\ }\href@noop {} {\bibfield  {journal} {\bibinfo  {journal} {Int. J. Multiphase Flow}\ }\textbf {\bibinfo {volume} {194}},\ \bibinfo {pages} {105419} (\bibinfo {year} {2026})}\BibitemShut {NoStop}%
\bibitem [{\citenamefont {Maeda}\ \emph {et~al.}(2021)\citenamefont {Maeda}, \citenamefont {Date}, \citenamefont {Sugiyama}, \citenamefont {Takagi},\ and\ \citenamefont {Matsumoto}}]{Maeda2021}%
  \BibitemOpen
  \bibfield  {author} {\bibinfo {author} {\bibfnamefont {K.}~\bibnamefont {Maeda}}, \bibinfo {author} {\bibfnamefont {M.}~\bibnamefont {Date}}, \bibinfo {author} {\bibfnamefont {K.}~\bibnamefont {Sugiyama}}, \bibinfo {author} {\bibfnamefont {S.}~\bibnamefont {Takagi}},\ and\ \bibinfo {author} {\bibfnamefont {Y.}~\bibnamefont {Matsumoto}},\ }\bibfield  {title} {\bibinfo {title} {Viscid--inviscid interactions of pairwise bubbles in a turbulent channel flow and their implications for bubble clustering},\ }\href@noop {} {\bibfield  {journal} {\bibinfo  {journal} {J. Fluid Mech.}\ }\textbf {\bibinfo {volume} {919}},\ \bibinfo {pages} {A30} (\bibinfo {year} {2021})}\BibitemShut {NoStop}%
\bibitem [{\citenamefont {Kusuno}\ and\ \citenamefont {Sanada}(2015)}]{Kusuno2015}%
  \BibitemOpen
  \bibfield  {author} {\bibinfo {author} {\bibfnamefont {H.}~\bibnamefont {Kusuno}}\ and\ \bibinfo {author} {\bibfnamefont {T.}~\bibnamefont {Sanada}},\ }\bibfield  {title} {\bibinfo {title} {Experimental investigation of the motion of a pair of bubbles at intermediate reynolds numbers},\ }\href@noop {} {\bibfield  {journal} {\bibinfo  {journal} {Multiphase Sci. Technol.}\ }\textbf {\bibinfo {volume} {27}},\ \bibinfo {pages} {51} (\bibinfo {year} {2015})}\BibitemShut {NoStop}%
\bibitem [{\citenamefont {Risso}(2018)}]{risso2018agitation}%
  \BibitemOpen
  \bibfield  {author} {\bibinfo {author} {\bibfnamefont {F.}~\bibnamefont {Risso}},\ }\bibfield  {title} {\bibinfo {title} {Agitation, mixing, and transfers induced by bubbles},\ }\href@noop {} {\bibfield  {journal} {\bibinfo  {journal} {Annu. Rev. Fluid Mech.}\ }\textbf {\bibinfo {volume} {50}},\ \bibinfo {pages} {25} (\bibinfo {year} {2018})}\BibitemShut {NoStop}%
\bibitem [{\citenamefont {Gong}\ \emph {et~al.}(2009)\citenamefont {Gong}, \citenamefont {Takagi},\ and\ \citenamefont {Matsumoto}}]{gong2009effect}%
  \BibitemOpen
  \bibfield  {author} {\bibinfo {author} {\bibfnamefont {X.}~\bibnamefont {Gong}}, \bibinfo {author} {\bibfnamefont {S.}~\bibnamefont {Takagi}},\ and\ \bibinfo {author} {\bibfnamefont {Y.}~\bibnamefont {Matsumoto}},\ }\bibfield  {title} {\bibinfo {title} {The effect of bubble-induced liquid flow on mass transfer in bubble plumes},\ }\href@noop {} {\bibfield  {journal} {\bibinfo  {journal} {Int. J. Multiphase Flow}\ }\textbf {\bibinfo {volume} {35}},\ \bibinfo {pages} {155} (\bibinfo {year} {2009})}\BibitemShut {NoStop}%
\bibitem [{\citenamefont {Wang}\ and\ \citenamefont {Socolofsky}(2015)}]{Wang2015}%
  \BibitemOpen
  \bibfield  {author} {\bibinfo {author} {\bibfnamefont {B.}~\bibnamefont {Wang}}\ and\ \bibinfo {author} {\bibfnamefont {S.~A.}\ \bibnamefont {Socolofsky}},\ }\bibfield  {title} {\bibinfo {title} {On the bubble rise velocity of a continually released bubble chain in still water and with crossflow},\ }\href@noop {} {\bibfield  {journal} {\bibinfo  {journal} {Phys. Fluids}\ }\textbf {\bibinfo {volume} {27}},\ \bibinfo {pages} {103301} (\bibinfo {year} {2015})}\BibitemShut {NoStop}%
\bibitem [{\citenamefont {Zhang}\ and\ \citenamefont {Ni}(2017)}]{Zhang2017}%
  \BibitemOpen
  \bibfield  {author} {\bibinfo {author} {\bibfnamefont {J.}~\bibnamefont {Zhang}}\ and\ \bibinfo {author} {\bibfnamefont {M.~J.}\ \bibnamefont {Ni}},\ }\bibfield  {title} {\bibinfo {title} {What happens to the vortex structures when the rising bubble transits from zigzag to spiral?},\ }\href@noop {} {\bibfield  {journal} {\bibinfo  {journal} {J. Fluid Mech.}\ }\textbf {\bibinfo {volume} {828}},\ \bibinfo {pages} {353} (\bibinfo {year} {2017})}\BibitemShut {NoStop}%
\bibitem [{\citenamefont {Zhang}\ \emph {et~al.}(2019)\citenamefont {Zhang}, \citenamefont {Chen},\ and\ \citenamefont {Ni}}]{Zhang2019}%
  \BibitemOpen
  \bibfield  {author} {\bibinfo {author} {\bibfnamefont {J.}~\bibnamefont {Zhang}}, \bibinfo {author} {\bibfnamefont {L.}~\bibnamefont {Chen}},\ and\ \bibinfo {author} {\bibfnamefont {M.~J.}\ \bibnamefont {Ni}},\ }\bibfield  {title} {\bibinfo {title} {Vortex interactions between a pair of bubbles rising side by side in ordinary viscous liquids},\ }\href@noop {} {\bibfield  {journal} {\bibinfo  {journal} {Phys. Rev. Fluids}\ }\textbf {\bibinfo {volume} {4}},\ \bibinfo {pages} {043604} (\bibinfo {year} {2019})}\BibitemShut {NoStop}%
\bibitem [{\citenamefont {Kusuno}\ and\ \citenamefont {Sanada}(2021)}]{Kusuno2021}%
  \BibitemOpen
  \bibfield  {author} {\bibinfo {author} {\bibfnamefont {H.}~\bibnamefont {Kusuno}}\ and\ \bibinfo {author} {\bibfnamefont {T.}~\bibnamefont {Sanada}},\ }\bibfield  {title} {\bibinfo {title} {Wake-induced lateral migration of approaching bubbles},\ }\href@noop {} {\bibfield  {journal} {\bibinfo  {journal} {Int. J. Multiphase Flow}\ }\textbf {\bibinfo {volume} {139}},\ \bibinfo {pages} {103639} (\bibinfo {year} {2021})}\BibitemShut {NoStop}%
\bibitem [{\citenamefont {Shirota}\ \emph {et~al.}(2008)\citenamefont {Shirota}, \citenamefont {Sanada}, \citenamefont {Sato},\ and\ \citenamefont {Watanabe}}]{Shirota2008}%
  \BibitemOpen
  \bibfield  {author} {\bibinfo {author} {\bibfnamefont {M.}~\bibnamefont {Shirota}}, \bibinfo {author} {\bibfnamefont {T.}~\bibnamefont {Sanada}}, \bibinfo {author} {\bibfnamefont {A.}~\bibnamefont {Sato}},\ and\ \bibinfo {author} {\bibfnamefont {M.}~\bibnamefont {Watanabe}},\ }\bibfield  {title} {\bibinfo {title} {Formation of a submillimeter bubble from an orifice using pulsed acoustic pressure waves in gas phase},\ }\href@noop {} {\bibfield  {journal} {\bibinfo  {journal} {Phys. Fluids}\ }\textbf {\bibinfo {volume} {20}},\ \bibinfo {pages} {043301} (\bibinfo {year} {2008})}\BibitemShut {NoStop}%
\bibitem [{\citenamefont {Takemura}\ \emph {et~al.}(2002)\citenamefont {Takemura}, \citenamefont {Takagi}, \citenamefont {Magnaudet},\ and\ \citenamefont {Matsumoto}}]{Takemura2002}%
  \BibitemOpen
  \bibfield  {author} {\bibinfo {author} {\bibfnamefont {F.}~\bibnamefont {Takemura}}, \bibinfo {author} {\bibfnamefont {S.}~\bibnamefont {Takagi}}, \bibinfo {author} {\bibfnamefont {J.}~\bibnamefont {Magnaudet}},\ and\ \bibinfo {author} {\bibfnamefont {Y.}~\bibnamefont {Matsumoto}},\ }\bibfield  {title} {\bibinfo {title} {Drag and lift forces on a bubble rising near a vertical wall in a viscous liquid},\ }\href@noop {} {\bibfield  {journal} {\bibinfo  {journal} {J. Fluid Mech.}\ }\textbf {\bibinfo {volume} {461}},\ \bibinfo {pages} {277} (\bibinfo {year} {2002})}\BibitemShut {NoStop}%
\bibitem [{\citenamefont {Clift}\ \emph {et~al.}(1978)\citenamefont {Clift}, \citenamefont {Grace},\ and\ \citenamefont {Weber}}]{clift1978bubbles}%
  \BibitemOpen
  \bibfield  {author} {\bibinfo {author} {\bibfnamefont {R.}~\bibnamefont {Clift}}, \bibinfo {author} {\bibfnamefont {J.~R.}\ \bibnamefont {Grace}},\ and\ \bibinfo {author} {\bibfnamefont {M.~E.}\ \bibnamefont {Weber}},\ }\href@noop {} {\emph {\bibinfo {title} {Bubbles, Drops, and Particles}}}\ (\bibinfo  {publisher} {Academic Press},\ \bibinfo {year} {1978})\BibitemShut {NoStop}%
\bibitem [{\citenamefont {Auton}(1987)}]{Auton1987}%
  \BibitemOpen
  \bibfield  {author} {\bibinfo {author} {\bibfnamefont {T.~R.}\ \bibnamefont {Auton}},\ }\bibfield  {title} {\bibinfo {title} {The lift force on a spherical body in a rotational flow},\ }\href@noop {} {\bibfield  {journal} {\bibinfo  {journal} {J. Fluid Mech.}\ }\textbf {\bibinfo {volume} {183}},\ \bibinfo {pages} {199} (\bibinfo {year} {1987})}\BibitemShut {NoStop}%
\bibitem [{\citenamefont {Ram{\'i}rez-Mu{\~n}oz}\ \emph {et~al.}(2011{\natexlab{a}})\citenamefont {Ram{\'i}rez-Mu{\~n}oz}, \citenamefont {Salinas-Rodr{\'i}guez}, \citenamefont {Soria},\ and\ \citenamefont {Gama-Goicochea}}]{RamirezMunoz2011a}%
  \BibitemOpen
  \bibfield  {author} {\bibinfo {author} {\bibfnamefont {J.}~\bibnamefont {Ram{\'i}rez-Mu{\~n}oz}}, \bibinfo {author} {\bibfnamefont {E.}~\bibnamefont {Salinas-Rodr{\'i}guez}}, \bibinfo {author} {\bibfnamefont {A.}~\bibnamefont {Soria}},\ and\ \bibinfo {author} {\bibfnamefont {A.}~\bibnamefont {Gama-Goicochea}},\ }\bibfield  {title} {\bibinfo {title} {Hydrodynamic interaction on large-reynolds-number aligned bubbles: Drag effects},\ }\href@noop {} {\bibfield  {journal} {\bibinfo  {journal} {Nucl. Eng. Des.}\ }\textbf {\bibinfo {volume} {241}},\ \bibinfo {pages} {2371} (\bibinfo {year} {2011}{\natexlab{a}})}\BibitemShut {NoStop}%
\bibitem [{\citenamefont {Ram{\'i}rez-Mu{\~n}oz}\ \emph {et~al.}(2011{\natexlab{b}})\citenamefont {Ram{\'i}rez-Mu{\~n}oz}, \citenamefont {Gama-Goicochea},\ and\ \citenamefont {Salinas-Rodr{\'i}guez}}]{RamirezMunoz2011b}%
  \BibitemOpen
  \bibfield  {author} {\bibinfo {author} {\bibfnamefont {J.}~\bibnamefont {Ram{\'i}rez-Mu{\~n}oz}}, \bibinfo {author} {\bibfnamefont {A.}~\bibnamefont {Gama-Goicochea}},\ and\ \bibinfo {author} {\bibfnamefont {E.}~\bibnamefont {Salinas-Rodr{\'i}guez}},\ }\bibfield  {title} {\bibinfo {title} {Drag force on interacting spherical bubbles rising in-line at large reynolds number},\ }\href@noop {} {\bibfield  {journal} {\bibinfo  {journal} {Int. J. Multiphase Flow}\ }\textbf {\bibinfo {volume} {37}},\ \bibinfo {pages} {983} (\bibinfo {year} {2011}{\natexlab{b}})}\BibitemShut {NoStop}%
\bibitem [{\citenamefont {Ram{\'i}rez-Mu{\~n}oz}\ \emph {et~al.}(2013)\citenamefont {Ram{\'i}rez-Mu{\~n}oz}, \citenamefont {Baz-Rodr{\'i}guez}, \citenamefont {Salinas-Rodr{\'i}guez}, \citenamefont {Castellanos-Sahag{\'u}n},\ and\ \citenamefont {Puebla}}]{RamirezMunoz2013}%
  \BibitemOpen
  \bibfield  {author} {\bibinfo {author} {\bibfnamefont {J.}~\bibnamefont {Ram{\'i}rez-Mu{\~n}oz}}, \bibinfo {author} {\bibfnamefont {S.}~\bibnamefont {Baz-Rodr{\'i}guez}}, \bibinfo {author} {\bibfnamefont {E.}~\bibnamefont {Salinas-Rodr{\'i}guez}}, \bibinfo {author} {\bibfnamefont {E.}~\bibnamefont {Castellanos-Sahag{\'u}n}},\ and\ \bibinfo {author} {\bibfnamefont {H.}~\bibnamefont {Puebla}},\ }\bibfield  {title} {\bibinfo {title} {Forces on aligned rising spherical bubbles at low-to-moderate reynolds number},\ }\href@noop {} {\bibfield  {journal} {\bibinfo  {journal} {Phys. Fluids}\ }\textbf {\bibinfo {volume} {25}},\ \bibinfo {pages} {093303} (\bibinfo {year} {2013})}\BibitemShut {NoStop}%
\bibitem [{\citenamefont {Zenit}\ and\ \citenamefont {Magnaudet}(2009)}]{zenit2009measurements}%
  \BibitemOpen
  \bibfield  {author} {\bibinfo {author} {\bibfnamefont {R.}~\bibnamefont {Zenit}}\ and\ \bibinfo {author} {\bibfnamefont {J.}~\bibnamefont {Magnaudet}},\ }\bibfield  {title} {\bibinfo {title} {Measurements of the streamwise vorticity in the wake of an oscillating bubble},\ }\href@noop {} {\bibfield  {journal} {\bibinfo  {journal} {Int. J. Multiphase Flow}\ }\textbf {\bibinfo {volume} {35}},\ \bibinfo {pages} {195} (\bibinfo {year} {2009})}\BibitemShut {NoStop}%
\end{thebibliography}%

\end{document}